\documentclass[sigconf]{acmart}
\usepackage{algorithm2e}
\usepackage{grffile} 
\usepackage{xcolor}
\usepackage{supertabular} 
\usepackage{cleveref}
\usepackage{multirow}
\usepackage{listings}
\usepackage{microtype}
\usepackage{newfloat}
\DeclareFloatingEnvironment[
   fileext=lol,
   name=Listing
]{listing}

\usepackage{natbib}

\usepackage{subcaption}
\DeclareCaptionSubType{listing}

\usepackage{mathtools} 
\newtheoremstyle{sltheorem}
{}                
{3 pt}             
{\slshape}        
{}                
{\bfseries}       
{.}               
{ }               
{}                
\theoremstyle{sltheorem}

\newtheorem{challenge}{Challenge}
\AfterEndEnvironment{challenge}{\noindent\ignorespaces}

\copyrightyear{2021} 
\acmYear{2021} 
\setcopyright{acmcopyright}\acmConference[ICAIF'21]{2nd ACM International Conference on AI in Finance}{November 3--5, 2021}{Virtual Event, USA}
\acmBooktitle{2nd ACM International Conference on AI in Finance (ICAIF'21), November 3--5, 2021, Virtual Event, USA}
\acmPrice{15.00}
\acmDOI{10.1145/3490354.3494408}
\acmISBN{978-1-4503-9148-1/21/11}

\usepackage{enumitem}
\usepackage{lipsum}
\setlist[itemize]{leftmargin=*}

\begin{document}

\title{On the Current and Emerging Challenges of Developing Fair and Ethical AI Solutions in Financial Services}

\author{Eren Kurshan}
\affiliation{
 \institution{Columbia University}
 \city{New York}
 \state{NY}
 \country{USA}
}
\email{ek2925@columbia.edu}

\author{Jiahao Chen}
\orcid{0000-0002-4357-6574}
\authornote{
Work done while at J.\ P.\ Morgan AI Research,
New York.
}
\affiliation{
 \institution{Parity}
 \city{New York}
 \state{NY}
 \country{USA}
}
\email{jiahao@getparity.ai}

\author{Victor Storchan}
\affiliation{
 \institution{J.\ P.\ Morgan}
 \city{Palo Alto}
 \state{CA}
 \country{USA}
}
\email{victor.storchan@jpmorgan.com}
\authornote{
\textbf{Disclaimer.}
\small{
This paper was prepared for informational purposes in part by J.P.Morgan Chase \& Co and its affiliates (``J.P. Morgan''), and is not a product of the Research Department of JP Morgan. JP Morgan makes no representation and warranty whatsoever and disclaims all liability, for the completeness, accuracy or reliability of the information contained herein.  This document is not intended as investment research or investment advice, or a recommendation, offer or solicitation for the purchase or sale of any security, financial instrument, financial product or service, or to be used in any way for evaluating the merits of participating in any transaction, and shall not constitute a solicitation under any jurisdiction or to any person, if such solicitation under such jurisdiction or to such person would be unlawful. © 2021 J.P.Morgan Chase \& Co. All rights reserved.
}
}

\author{Hongda Shen}
\affiliation{
  \institution{Univ. Alabama}
  \city{Huntsville}
  \state{AL}
  \country{USA}
  }
\email{hs0017@alumni.uah.edu}

%


\begin{abstract}
AI has found a wide range of application areas in the financial services industry. As the number and the criticality of the applications continue to increase, fair and ethical AI has emerged as an industry-wide objective. In recent years, numerous ethical principles, guidelines and techniques have been proposed. However, the model development organizations face serious challenges in building ethical AI solutions. This paper focuses on the overarching issues model development teams face, which range from the design and implementation complexities, to the shortage of tools, and the lack of organizational constructs. It argues that focusing on the practical considerations is an important step in bridging the gap between the high-level ethics principles and the deployed AI applications, as well as starting industry-wide conversations toward solution approaches. 
\end{abstract}

\settopmatter{printfolios=true} 
\maketitle



\section{Introduction}
\label{sec:intro}

Over the last decade, AI has gained significant interest in the financial services industry. It has been used in all key business functions from operations to new product development, customer service, marketing and risk management. According to the latest industry reports, AI implementation efforts across the core business functions is around ~90\% in the leading financial firms \citep{wef2020}. Today, financial services functions like market making, trading, investment management, credit services and retail banking increasingly rely on AI \citep{Australia}.  Due to its growing prominence in finance, building fair and ethical AI solutions has become an industry-wide goal.  

\noindent Industry surveys indicate that over 50\% of the financial firms perceive AI ethics as a major or extreme concern, yet only a third report preparedness for potential issues \citep{Deloitte_AI}.  These concerns are quite valid as numerous studies showed that AI is capable of exhibiting unethical behaviors. AI agents have been shown to produce biased outcomes \citep{microsoft2,google} and even demonstrate the ability to independently develop deceptive tactics such as deception in experimental settings \citep{deception,lying}. Although a large number of techniques have been developed in recent years, the practice of ethical AI techniques in finance is still in its nascent stages.   

\noindent To date, ethical AI practice has been largely focused on the regulatory mandates on fairness and explainability in financial services.  Model development organizations face a wide range of issues in building ethical applications ranging from organizational issues to the lack of standards and methodologies.  This paper overviews the existing and emerging complexities of the ethical AI practice and argues that addressing the broader challenges is critically important in ensuring ethical behavior. The purpose of this paper is not to provide a comprehensive overview of the current practices or techniques (as it should be understood that they change relatively quickly). The paper has 2 main goals, as it: (i) aims to highlight overarching challenges and pain-points of building ethical solutions for financial services teams; (ii) tries to start industry-wide conversations across the organizational lines to tackle the pervasive, systematic and organizational challenges.

\noindent The paper is structured as follows: 
\Cref{sec:needforethics} discusses the need for ethical AI; \Cref{sec:challenges} overviews the practical issues in ethical AI development from a model development perspective;
and finally \Cref{sec:conclusion} outlines potential solution opportunities and outlook.

\section{The Need for Ethical AI in Finance}
\label{sec:needforethics}



Financial services firms show significant diversity in their practices of AI and AI ethics. Therefore, this section aims to present selected highlights instead of a comprehensive overview. 

\noindent\textit{\textbf{Traditional Financial Service Providers:}}
(i) \textit{Credit Scoring and Credit Limits:} Adjusting customers credit risk evaluations and scores based on affinity profiling has been a common theme in the  FTC complaints \citep{amex,FTC}. Similarly, gender-based discrimination has been reported in the credit limit decisions in the emerging payment systems \citep{applecard}. (ii)\textit{ Lending:} In the U.S., credit models are expected to comply with numerous regulations enforced at local, federal and state levels (such as ECOA).  A wide range of fairness, bias and discrimination issues have been reported in lending applications as regulatory violations according to the CFPB \citep{CFPB}. (iv) \textit{ Operational Models:}  Financial firms use AI pervasively for their operations ranging from customer service, to  customer loyalty programs, and authentication etc., each of which have varying levels of ethical risks.  Image processing (used in authentication solutions, security functions etc.) have been shown to carry bias risk \citep{MIT}.  Similarly, NLP (used in the customer service applications, financial assistants, recruiting, and personnel management) has a tendency to inherit and exhibit discriminatory patterns \citep{ferrer,NLP_bias,FINRA}, even when preventative measures are taken. Overall, there has been limited focus on the ethical profiles of the operational models in financial services.  

\noindent\textit{\textbf{FinTech \& Big Tech:}}
Financial technology firms experienced a remarkable growth in the past few years. According to the industry surveys, fintech has reached over 60\% global adoption rate \citep{EYFintech}. Consequently, the reports of bias and other ethical issues also increased in recent years \citep{Gulam}. As non-bank lenders face less regulatory scrutiny, their AI models pose growing ethical concerns \citep{DarkSide}, among which unorthodox data usage and feature selection emerge as common problem areas \citep{FTC}. According to \citep{berkeley2}, global fintech firms utilize thousands of features for credit scoring and lending. Digital footprints are frequently used to analyze the individuals in many dimensions and serve as a gateway for protected class information (like age, race, gender, marital status, disability etc.) \citep{india2}.  The use of customers marital and dating status, social media profiles, SMS message contents \citep{india1}, activity tracker data \citep{capone}, typing speed and accuracy, customers life-stage (e.g., growing family vs. empty nester) \citep{penn}, facial analysis and micro-expressions have been reported for lending decisions \citep{wsj2}. These create concerns related to privacy, transparency, autonomy and fairness.  The limited regulatory oversight, along with fintechs increased prominence in the finance exacerbate the impact of the underlying ethical issues \citep{DarkSide}. In recent years, technology companies have been entering the financial services industry directly or through partnership with financial firms \citep{kpmg}. These trends further motivated the recent debates for more regulatory oversight on technology companies. 

\vspace{8pt}
\section{Practical Challenges of Ethical AI }
\label{sec:challenges}


\subsection{Conceptual Challenges}
\begin{challenge}
  The complexity and the interdisciplinary nature of ethics is perceived as an obstacle for many development organizations. 
\end{challenge}

\noindent\textit{\textbf{Complexity:}}  Despite the emerging interest, dedicated AI ethics efforts are still not common among the model development teams in financial services. 
Ethics involves numerous interdisciplinary theories, hard to quantify criteria, and remarkable breadth and depth. These factors make it difficult for the development organizations to come up with clear goals and strategies \citep{AAAI_Anderson}. Hence, they often perceive this complexity as an obstacle in starting AI ethics efforts.  Defining and implementing ethical AI behavior for a wide range of financial AI applications (from trading to retail banking applications) is a non-trivial task.

\noindent\textit{\textbf{Scope:}} (i) \textit{Long list of Principles:} During the last few years, a large number of AI ethics guidelines and policy documents has been published globally, each presenting a long list of principles.   Recent studies have quantitatively analyzed these documents to pinpoint the shared principles \citep{ETH}. Even from a limited set of guidelines, over 60 individual principles have been identified \citep{ETH}. Out of these, the top 5 shared principles were: \textit{transparency} (encompassing explainability, interpretability etc.), \textit{fairness} (including but not limited to non-bias, non-discrimination, equality), \textit{non-maleficence} (capturing the primum non nocere, security, safety), \textit{responsibility} (consisting of accountability, liability etc.) and \textit{privacy}. Guideline documents exhibit significant divergence with respect to their specificity, as well as the interpretation of the principles and their implementations. This also yields consequential variations in their overall impact \citep{Guides}.  (ii) \textit{Limited Application Span:} Banks tend to focus on the regulated and higher-visibility models for ethical AI, and rely on the legal and regulatory requirements as the primary guidelines. However,
the reports of ethical issues in unexpected applications and unregulated industries showcase that this limited scope carries significant risks. (iii) \textit{Fintech:} The limited regulatory oversight of the financial technology firms also poses a significant risk as noted earlier \citep{DarkSide}.

\begin{challenge}
Ethical norms are not universal or static, showing variations over time, across applications, and cultures.
\end{challenge}

\noindent  The inherent variations among ethical norms have been showcased by various studies: (i) \textit{Context and Application-Specific Variations:} Acceptable norms in one application or context may not be acceptable in others \citep{selbst}. (ii) \textit{Population Variations:} Ethical variations are common across populations,  geographical regions and cultures \citep{awad},\citep{Kahane1}. (iii) \textit{Temporal Variations:} Ethical norms change over time and in some cases relatively quickly. Examples of such changes can be found in the medical ethics \citep{Calman}(such as the shift from beneficence to the increased patient autonomy in medicine \citep{Bioethics}). Despite the need to fine-tune the AI-based solutions for such variations, relying on \textit{moral subjectivity} carries tremendous risks for AI \citep{EmotionalMoral}. Graded governance frameworks were proposed to capture the variations at population, culture and individual-level \citep{WEF}. However, current AI applications do not have access to such frameworks.

\subsection{Design-Time Considerations}
\begin{challenge}
Picking the right approach from a rapidly expanding list of fairness techniques is a difficult task for the development teams. 
\end{challenge}

\noindent \textit{\textbf{Fast-Growing List of Criteria:}} Even in the past few years alone, a large number of fairness definitions and metrics has been proposed \citep{madras,roth,mehrabi}. This rapidly-expanding list and the complexity of the individual criterion are challenging from an implementation perspective \citep{harvard,MIT_bias}. Moreover, in the past few years, studies have started exploring the inherent limitations, conflicts, interactions and the unexpected long-term consequences of the state-of-the-art techniques \citep{google_gym,berk,Cambridge}, which results in open-ended and difficult explorations for the development teams.

\noindent \textit{\textbf{Process Complexity:}} (i) \textit{Model Governance Timelines:} As a part of the model risk management review, AI teams are expected to justify all design decisions made during the model development \citep{ICAIF}. This implies that the selected fairness criteria has to be compared with all viable alternatives and justified for its own merits. The extensive list of techniques, combined with the lack of industry standards, increases the amount of time required to complete the analysis and reviews. As a result, some development teams try to avoid using AI ethics techniques if there is no clear regulatory mandate.  (ii) \textit{Design-time Analysis:} Likewise,  model development organizations are frequently overwhelmed with the difficulties of picking and implementing the right set of techniques during development. Since there are no standards, the effort to incorporate the fairness metrics scales with the number of alternative approaches.

\begin{challenge}
 Ethics principles involve complex design trade-offs, with no clear guidelines on how to manage them.
\end{challenge}

\noindent \textit{\textbf{Design Trade-offs:}}
During the development process, even a few principles translate to complex, multi-objective optimization problems for AI \citep{ETH} (involving numerous design trade-offs and possible conflicts \citep{Cambridge}). Navigating this complex design space requires tool and infrastructure support in addition to consistent and clear criteria.  Yet, AI  development organizations have limited to no access to such tools, supporting infrastructures or guidelines.  

\noindent\textit{\textbf{Incompatible or Conflicting Criteria:}} The difficulties and in some cases the potential impossibility of satisfying multiple principles at the same time have been demonstrated by studies like \citep{impossibilityoffairness}. (i) \textit{Explainability vs. Accuracy:} The possible clashes between explainability and model accuracy have been studied by many researchers in the machine learning community \citep{ExplainabilityvsAccuracy}. Though new studies posit that there is no inherent conflict between the two principles \citep{Rudin}, explainability remains to be challenging for most high-performance machine learning approaches (like deep learning) \citep{DARPA}. (ii) \textit{Privacy vs. Fairness:} Potential compatibility of fairness and privacy principles raise fundamental questions \citep{Ekstrand}.  \citep{privacy} claims that it is impossible to achieve exact fairness with differential privacy. The resulting conflicts showcase the need for more guidance during the design and development stages. (iii) \textit{Fairness Criteria:} Conflicts can also occur among the criteria used for a single ethics principle such as fairness. \citep{berk,kleinberg} argue that it is not possible to simultaneously satisfy calibration and forms of classification parity for fairness. Similarly, \citep{Chouldechova} shows the incompatibility of the equal positive/negative predictive values, and the equal false positive/negative rates if base rates differ across groups.



\noindent \textit{\textbf{Transparency vs. Security and Privacy:}}
Transparency is a critically important tool to ensure trustworthy AI behavior. The black-box nature of AI in commercial and unregulated applications has been a growing concern in the industry \citep{EthicsofAIEthics}.  Intellectual property protections and liability considerations appear to play a significant role in the secrecy.  Algorithmic transparency aims to make the underlying algorithms, and in some cases the model data, public. However, it faces confidentiality and security issues in financial use cases. A large number of models (such as transaction processing, fraud detection, risk models) lose effectiveness with even limited public information sharing. In transaction processing and retail banking applications, criminals quickly develop custom attacks to bypass the fraud detection models. Novel solution approaches are needed to balance these opposing forces and enable increased transparency without sacrificing the security and privacy.

\begin{challenge}
\vspace{-5pt}
The selection of fairness definitions and metrics is a complex process, requiring analysis to ensure long-term consequences, compatibility with other ethics and fairness principles etc.
\end{challenge}

\noindent\textit{\textbf{Long-Term Consequences and Sufficiency:}} Fairness techniques may have different effects over time. A technique that helps the protected classes in the short-term may end up hurting them in the long-term. These effects are hard to predict and are often counterintuitive \citep{google_gym}. They require tool and infrastructure support as well as extensive analysis. Others studies raised important questions on the quality and the adequacy of the existing techniques in detecting bias and discrimination \citep{goel}. 

\noindent\textit{\textbf{External Interactions:}} AI applications in finance are often tightly intertwined with various external processes. As AI-based decisions have growing influence in the society, individuals have the incentive to change their behavior to positively affect the outcomes \citep{hardt18}. This interconnectivity often yields unexpected interactions and occasionally positive feedback loops. Identifying, analyzing and profiling external interaction and feedback patterns are essential, due to their potential to go out of control and yield negative outcomes \citep{goel,barocas}. However, these efforts are highly demanding and require significant research investments.

\noindent \textit{\textbf{Counterintuitive Effects:}} In credit decisions, some scholars emphasize counterfactual fairness (i.e. the outcome would not change if the entity belonged to a different class) \citep{kusner}, while others assert that AI systems should have the capacity to compensate for the existing social inequalities.  Recently, it has been shown that compensation techniques have the potential to negatively impact the long-term outcomes for the protected classes \citep{Hu,hardt18}, which motivates thorough analysis of the consequences of such approaches during design and development stages.



\subsection{Implementation Challenges}

\begin{challenge}
Ethical principles and processes are often hard to quantify and implement in AI applications.
\end{challenge}
\noindent
\textit{\textbf{Individual Principles \& Criteria:}}Fairness definitions have been quite effective in providing metrics and criteria for AI applications in finance \citep{Narayanan}. However, beyond fairness, ethical principles are harder to quantify and implement in computing systems. A multitude of context-specific intricacies, application-specific variations, singularities, conflicts and boundary cases make the practical implementations complex and the resulting  characteristics unclear \citep{anderson}.
\textit{\textbf{Processes \& Systems:}} Focusing solely on the individual principles is not sufficient for ethical behavior, as the supporting architectures and mechanisms to operationalize the principles are missing. As an example, humans prioritize among different principles and rules (of varying rigidity and importance) while making ethical judgements \citep{Bartels}. Presently, AI lacks such mechanisms and the system architecture to accommodate ethical processes.

\begin{challenge}
 Almost all stages of the AI development process impact the resulting ethical profile.
\end{challenge}

\noindent\textit{\textbf{Many Sources of Bias:}} Model development involves numerous stages, each of which can affect the overall behavior of the model \citep{barocas}. (i)\textit{ Problem Formulation:} Bias may be introduced as early as the initial problem formulation and during the high-level definition of the solution \citep{Solon}.  (ii) \textit{Data Biases:} may occur during the data source selection, data acquisition, analysis, evaluation, measurement, sampling, aggregation and representation. Biases during the predevelopment include historical data biases, temporal bias, external biases, historical interaction and channel biases, data limitation biases, data quality and processing biases. (iii) \textit{Algorithmic Bias:}  Later, during the algorithm selection, solution design, algorithm selection, objective function selection, and threshold selection stages need to be carefully designed for prevention. During the feature selection, each feature candidate carries the potential for bias. The omitted variables and feature preprocessing may also inject bias. \citep{hooker2020characterising} shows that compressed models can also over amplify the bias compared to a non compressed baseline. (iv) \textit{Interaction Bias:} During the model execution, bias may occur as a result of the interaction differences between segments (such as the mobile banking application usage among young and elderly client populations). Financial applications crucially need specialized techniques for data quality, data hygiene, provenance and debiasing to address the underlying data problems: (i) \textit{ Debiasing \& Detection:} Specialized techniques and processes are needed to detect, assess and mitigate bias; (ii)\textit{ Traceability \& Provenance:} Required to track and manage the model data throughout its life-cycle; (ii)\textit{ Data Quality Criteria:} Such as metrics, benchmarks, and data quality frameworks \citep{DataQuality}); (iii)\textit{ Tool \& Infrastructure Support:} Needed to cover the end-to-end model life-cycle for bias detection and mitigation.

\noindent\textit{\textbf{Data \& Model Ownership:}} In most financial firms, the data used for model development comes from numerous sources, owned and maintained by different organizations and lines-of-businesses (LOBs). Yet, current AI ethics efforts and mandates remain mostly within the development organizations. AI teams usually have limited/no access to the source databases and reference data systems. Furthermore, they have limited control over the data during its life-cycle (including the upstream data processes until the data reaches the model development systems). Firm-wide data ethics processes and organizational changes are needed to ensure that the model data is acquired, processed and used according to the firms AI standards and guidelines.

\begin{challenge}
Model development alone does not guarantee ethical behavior; post-development and deployment are equally important.

\end{challenge}

\noindent (i)\textit{ Testing and Profiling:} Due to its inherent opacity, AI requires extensive profiling to assess its behavior in different scenarios. Currently, AI testing and profiling stages are primarily focused on the model performance and compliance. In order to achieve the desired deployment behavior, AI applications need to be thoroughly profiled and tested for their adherence to the ethics principles in different scenarios.  (ii)\textit{ Post-Development Stages:} AI ethics procedures need to cover the entire model life-cycle to be effective. Post-development stages including model validation, model maintenance, monitoring and audit are essential to ensure the desired behavioral profile.  (iii)\textit{Monitoring \& Mitigation:} Currently, regulated models are go through on-going monitoring reviews periodically (with minimum of once a year). Beyond the periodic reviews, AI and machine learning models are rarely continuously monitored in real-time \citep{ICAIF}. As a result, it is hard to detect and mitigate issues in a timely manner. \citep{darkrobot} uses robotics examples to illustrate the dangers of solely focusing on the development stages for ethics. Without real-time mitigation,  models may be exposed to interventions and change their behavior over time. Ethics architectures and design-layers have been explored to further ensure consistent and reliable behavior during run-time \citep{ethicslayer,bottomup}. With AI's ever-growing application list and sophistication, similar constructs may be needed for financial applications in the near future. (iv)\textit{ Socio-Technical Systems:} \citep{10.1145/3313831.3376219,tsirtsis2020decisions} argues factors such as trust, algorithmic recourse and  human-interactions are key components to maximize the overall deployment system utility.

\begin{challenge} 
Ethical AI is hard to define and implement in the broader financial applications.
\end{challenge}

\noindent\textit{\textbf{Defining What is Ethical:}} Even though AI has been pervasively used across the industry, defining what is ethical has been considered a difficult task for a wide range of financial applications (including trading systems, financial advisory services, investment and portfolio management systems, customer service applications, marketing and loyalty campaigns, retail banking applications etc). \citep{wallach} discusses the potential dangers of AI through various scenarios, including an AI-based trading system that makes decisions with no ethical considerations or capabilities . Such systems have the potential to cause remarkably negative outcomes without concern for the broader economic, societal or ethical implications of their decisions. However, so far, there has been very limited work on the definition of ethical behavior for the broad range financial applications that use AI (with the exception of a few areas like high-frequency trading algorithms \citep{HFTEthics}). Similarly, the implications of ethics principles (beyond fairness and explainability) need to be considered for the full application scope. For example, autonomy is an important ethics goal according to numerous guidelines, yet the implications on customer interaction models (marketing, sales and user experience systems) or retail banking applications of AI has not been analyzed.  

\noindent\textit{\textbf{Application \& Scale:}} 
(i) \textit{Unforeseen Issues in Unexpected Applications:} A large portion of the AI ethics issues has been reported in unexpected applications like chatbots, image labeling and face recognition systems. The limited purview of the ethical AI practices in finance is a significant drawback in addressing such unforeseen issues. This underscores the need to incorporate the design principles pervasively and universally across all AI deployments. (ii) \textit{Scale of Impact:} In other application areas, AI models have been used to build economies of attention like social media. In these systems attention is harvested and traded whether or not the attention-seeking behavior is moral \citep{guardian2}.  By nature, the financial services industry also has significant economic and societal impact, in which the scale of AI's impact is currently not accounted for \citep{AIImpact}. Designing AI applications for their full potential instead of the incremental value-add may yield improved outcomes in terms of such broader impact.

\noindent\textit{\textbf{Beyond Individual Models:}} 
(i) \textit{Individual Models vs. Complex Systems:} In most financial firms, decision systems involve numerous interconnected and interdependent models. However, the definition of ethical behavior at the individual model-level vs. system-level remains unclear for most systems.  (ii) \textit{External Interactions \& Feedback Loops:} Algorithmic solutions often exhibit strong interactions with the external environment \citep{goel}. In 2013, algorithmic trading systems spiraled out of control (by going into a positive feedback loop on shorting) when the Associated Press's Twitter account was hacked \citep{Karppi}. Such robustness problems may have ethical implications for other AI applications in the future.  Without AI ethics practices and run-time oversight, autonomous AI agents may adopt fraudulent behaviors like bid-rigging, price fixing and other forms of collusion \citep{Collusion}.

\vspace{-5pt}

\subsection{Guidelines, Standards and Regulations}

\begin{challenge}
Development organizations face a gap between the high-level ethics guidelines and the deployment systems.
\end{challenge}

\noindent\textit{\textbf{Principles \& Guidelines:}}  In recent years, a large number of AI ethics guideline documents have been published: International guidelines (e.g. OECD Principles \citep{OECD}, E.U. Guidelines, G7 Vision on AI), country guidelines (e.g. U.K., Singapore, Australia), association guidelines (e.g. ACM \citep{acm}, IEEE Principles \citep{IEEE_Principles}), AI non-profit and consortium guidelines (e.g. Asilomar Principles \citep{asilomar}), company guidelines (e.g. Google, Microsoft, IBM), state and regional guidelines etc.  These policy and guideline documents include a long list of high-level principles and recommendations, while providing limited guidance on specifics, interpretation, accountability and implementation aspects \citep{ETH}, which affect their impact factor according to studies like \citep{AIImpact}. Some scholars have criticised and questioned the overall adequacy and sufficiency \citep{EthicsofAIEthics}. However, principles and high-level guidelines had limited impact on the practice so far \citep{Principles}. 

\noindent\textit{\textbf{Application \& Industry-wide Guidelines:}}  While high-level guidelines are valuable, a full-stack of supporting rules and processes is necessary to make them actionable during the implementation: (i) \textit{ Application-Specific Guidelines:}  Application-specific guidelines are needed to ensure consistent AI behavior across in financial applications such as mortgages, lending, credit scoring, financial advisors, risk management etc. (ii)  \textit{Industry Standards:} Financial industry AI standards may provide critical guidance, similar to the robotics standards like BSIs BS8611, IEEEs P7000 that outline the potential risk categories and guidance on the mitigation strategies. (iii) \textit{Administrative Organizations and Codes of Practice:} Codes of ethics for AI, similar to the code of ethics for medical professionals, have been proposed by some researchers \citep{Veliz19,Stanford_AI}. Likewise, social license concepts \citep{DeloitteDL} and specialized regulatory entities (similar to FDA and other regulators) have been suggested for industry-wide guidance \citep{FDA}. Regardless of the alternative paths, the need for clear and thorough guidelines is evident.

\noindent\textit{\textbf{Firm-wide Policies, Standards \& Processes:}} (i) \textit{Ethics Policies and Frameworks:} Similar to the existing risk management frameworks and risk policies in the financial institutions, firm-wide ethics frameworks and ethics policies are essential in ensuring the desired outcomes as well as consistent implementation practices across the firm. Since the resulting AI behavior is impacted by all design stages, end-to-end coverage of the model life-cycle (from the problem formulation, to development, testing,  run-time monitoring and mitigation) is imperative to achieve this. Furthermore, extending the European Commission's traceability criteria for AI is critical in the highly-complex model development landscape (which requires traceability for all datasets and processes involved in the data gathering and labelling phases) \citep{EUTraceability}. (ii) \textit{Standards and Metrics:} Additionally, firm-wide and application-specific (quantitative and qualitative) criteria are required to standardize the firm-wide development efforts. Post-development, similar criteria can be used for run-time management and mitigation purposes (such as alerting and mitigation mechanisms). (iii) \textit{Libraries:} Firm-wide libraries are essential for the pervasive dissemination and implementation of the ethical AI principles, rules and guidelines. Libraries may include references for the principles, quantitative and qualitative criteria, process guidelines, ethics review guidelines, application-specific standards, operational specifications, standard ethics tools and techniques etc.

\noindent \textit{\textbf{Tools \& Methodologies:}}
As a result of the infrastructure and tool limitations, AI development teams are frequently compelled to pick one approach against another without the ability to perform thorough analysis to support their decisions. In the last few years, a number of AI ethics tools have been presented \citep{Fairness360,whatif}, yet many external AI tools are not permissible or approved for internal use in the financial firms.  Thus, financial AI models are in need of industry-grade and internally approved tool and infrastructure support to improve the resulting model behavior.

\noindent \textit{\textbf{Regulatory Guidance:}}
Regulations are unique in their ability to enforce the desired behavior (at times through fines and penalties), as well as reducing the complexity and subjectivity of the ethical criteria. (i) \textit{Fine-tuning Existing Regulations:} According to the industry reports, most financial regulators are in the process of reviewing the existing financial regulations for potential adjustments and expansions to address the growing use of AI \citep{Deloitte,OCC}. In 2019, SEC announced rule amendments to modernize the risk factor disclosures that are mandated pursuant to the Regulation S-K, to capture the emerging concerns on AI risks through materiality \citep{SEC}. According to the news reports, FDIC has been pushing for expanded AI regulations \citep{FDIC}. CFTC and FINRA published their comprehensive reports with recommendations on AI use \citep{FINRA,CFTC}. More regulatory guidance for AI and AI ethics is expected in the next few years.  (ii) \textit{New Regulations:} New regulations were introduced in response to the latest AI and data concerns.  These regulations have strong emphasis on transparency, accountability, interpretability and explainability, while their characteristics range from broad-scope to local and application-specific acts (e.g. the Illinois Act on the use of AI in specific applications \citep{illinois}). (iii) \textit{Next Generation Regulations:} Lately, there has been significant interest in regulation development for AI (such as the OECD efforts to support a global governance framework \citep{OECD}). Numerous regulators, including the Basel Committee, held workshops towards future guidance for AI \citep{Basel2} (akin to earlier data completeness and data quality efforts). The upcoming EU AI Act requires the financial firms to fulfill conformity assessments and categorizes most financial services (e.g. credit scoring system, underwriting tools) as high-risk \citep{aireg}. (iv)\textit{Regulatory Enforcement:}  Regulatory enforcement is an important component of ensuring ethical behavior. Exploring the limitations of the current enforcement practices and the use of regulatory technology to assist enforcement \citep{DarkSide} have gained significant interest lately. The operational aspect of some of the upcoming E.U. regulations \citep{Veale} require clarifications and further work for enforcement.



\noindent \textit{\textbf{Human Judgement:}} 
Although they remain controversial, numerous tests have been proposed to capture the human moral judgement. (i) \textit{Human Judgement Tests:}  Comparative Moral Turing Tests (cMTT) rely on humans to judge the ethical performance of humans and machines \citep{cmtt}, Ethical Turing Tests, use committees of ethicists to make decisions on the agreeableness of the AI decisions \citep{ethical_turing}.  (ii) \textit{Human-in-the-loop Decision-making:} Human-in-the-loop decisions have been explored in various industries to improve the trustworthiness of AI.  Due to the prominence of the high-bandwidth and real-time applications in the financial services industry, such approaches may be confined to a limited set of applications.  Furthermore, as noted in \citep{Accountability} 
both humans and AI have their unique strengths and weaknesses in ethical judgement. Designing for such strengths and weaknesses is essential for solution design. (iii) \textit{Reviews and AI Ethics Committees:} Reviews and executive-in-charge models have been proposed for AI ethics evaluations \citep{FinanceAI}. Also, in the past few years, some companies have announced firm-wide AI ethics committees. Despite the on-going interest, at this point in time, the adoption rate has been lower in the financial firms and the reports of dismantling and overruling of AI committees in other industries illustrate the uncertainty of the outlook \citep{Accenture_AI_Committee,BBC,Financialtimes}.
\vspace{-15pt}
\subsection{Process \& Organizational Considerations}

\begin{challenge}
Contrary to the popular belief, AI development organizations have very limited power in the modeling decisions.
\end{challenge}
Despite the external perception, in financial firms, most of the design decisions are controlled by the business units and the stakeholders. Model development and technical organizations are generally in service roles, responsible for implementing the decisions. One of the practical challenges is that decision making groups (like the business teams and stakeholders) frequently have limited/no understanding of the technical or ethical considerations of AI. The inherent complexity of AI ethics makes the decision-making process and the final solutions prone to issues. Despite their limited power, the model development organizations are still key, as intermediaries in ensuring that AI ethics decisions are made across the organizational boundaries. 

\begin{challenge}
 Although AI is pervasively used in almost all of its functions, financial services industry lacks organizational constructs and processes for ethical and fair AI.
\end{challenge}

\noindent \textit{\textbf{Organizational Boundaries \& Role Definitions:}}
(i) \textit{System vs. Model Ownership:} As noted earlier, AI is frequently a component in much larger and complex computational systems. AI components within such systems are frequently developed and owned by different teams and LOBs. As a result of this distributed and complex ownership practice, it is hard to achieve the desired system-level characteristics by solely focusing on the individual AI components. (ii) \textit{Data Ownership:} Akin to the model ownership concerns, reference data systems are typically owned and managed by number of different organizations. This leaves gaps in the overall design flow, which cannot be solely resolved by the efforts of the model development teams, as they have limited/no-access to the reference data sources or the processes being used. (iii) \textit{Role Definitions:} AI development involves a large number of developers as well as involvement from business units, data owners, risk and compliance teams, deployment teams etc. Currently there is no clear guidance on the role definitions for the participating teams (or ownership for ethics-related decisions). Most of the AI ethics efforts are voluntarily driven the model development organizations, yet they have limited power in the overall process.

\noindent \textit{\textbf{AI Ethics Organizations:}}  Even though firm-level ethics committees have been assembled by a number of companies, relying on ethics committees alone leaves large operational gaps in the AI design and development process. Building dedicated model ethics organizations (with direct involvement and responsibility in the model development) is key in closing these gaps. AI ethics organizations may fit within the existing model risk management framework, with or within the MRM organizational constructs. Despite the well-publicized diversity challenges in the AI development organizations \citep{GenderRace}, the decision-making groups (such as stakeholders, line-of-business leads) frequently exhibit much more prominent diversity issues. Hence, it is not possible to represent the subtleties of human judgement with the current development and decision-maker populations. Ethics committees and AI ethics organizations need to be formed cognizant of such difficulties.

\noindent \textit{\textbf{Resource \& Time Pressures:}} (i) \textit{Development Constraints:} The demand for AI and machine learning development has been surging in the financial services industry. According to the industry surveys, the number of models in the large financial firms increase by \textit{10-25\%} each year \citep{mckinsey_modelcount}. Model development teams face serious resource and time pressures, leaving limited bandwidth for AI ethics initiatives. (ii) \textit{Model Risk Management Constraints:} Comparably, model risk management groups face resource and bandwidth constraints, which affects the AI development and governance timelines negatively. Model governance process was designed for traditional financial models like capital analysis and faces problems in assessing and managing AI risk. Risk management reviews frequently take many months to sometimes years \citep{ICAIF} (even without AI ethics techniques).  For many development teams, this creates pressures to avoid sophisticated AI ethics techniques to meet the governance review timelines. The lack of clear ethics guidelines and theoretical guarantees on AI performance, makes the model governance for AI ethics challenging and time consuming.




\noindent \textit{\textbf{Community Involvement:}} (i) \textit{Industry Community:}  As a result of the organizational restrictions, most AI development teams in the financial firms have limited involvement in the broader AI community, which makes it difficult to follow and adopt the state-of-the-art techniques. (ii) \textit{Academic Community:} Likewise, the limited data sharing between the industry and academic groups often reduces the applicability of the resulting academic techniques in  deployment systems. The highly-sensitive nature of the financial data and the resulting access restrictions are major obstacles in this process.  Latest proposals attempt to tackle these issues by improving the quality of synthetic data \citep{JPM} and building legal-technical frameworks in order to balance the opposing openness and privacy forces \citep{Open}.  (iii) \textit{Non-Profits and AI Consortia:} Lately, a number of non-profits and AI consortia has been formed to primarily focus on responsible and trustworthy AI development (e.g. Partnership on AI, Machine Intelligence Research Institute). These efforts play an important role in trying bridge the gap between academia, industry, governments and regulators.

\subsection{Emerging Challenges \& Autonomy}
\label{sec:Emerging}

\begin{challenge}
 Ensuring ethical behavior in autonomous agents is significantly more difficult than the current AI systems.
\end{challenge}
(i) \textit{Increased Autonomy:} Adaptive machine learning approaches have gained significant interest in recent years (e.g. online and reinforcement learning) \citep{MnihKSR2015,XiongLZY2018}. Despite their performance advantages, adaptive techniques like reinforcement models are known to surprise developers with their unexpected actions \citep{ieee}, which cause issues in ensuring ethical behavior \citep{unpredictableAI}. (ii) \textit{Enhanced Ethical Agency:} AI is pervasively used in increasingly sophisticated decision systems for risk management, trading, financial market forecasting, portfolio management \citep{portfolio}.  Implicitly ethical AI has limited capacity to perform ethical scenario analysis or trade-offs. However, with the growing sophistication of AI applications and increased autonomy, these capabilities may be needed soon. AI may be faced with limited and rigid modes of operation and undesired outcomes without such capabilities. Scenario-based ethical analysis has been studied for robotics use cases in controlled lab environments \citep{darkrobot}, and remains to be challenging outside of such structured and supervised set-ups. 


\vspace{-3pt}

\begin{challenge}
AI may exhibit novel unethical behaviors and discrimination patterns, which may be hard to discover and mitigate. 
\end{challenge}
\noindent \textit{\textbf{Novel Issues:}} As noted in \Cref{sec:intro}, AI systems have demonstrated the ability to learn and independently develop unethical strategies  (e.g. deception \citep{deception}, aggression \citep{aggression}). In the near future, it will be vital to assess AI's novel behavioral patterns from an ethical perspective. This illustrates the importance of: (i) \textit{Extending the Scope of AI Ethics Practice:} Broadening the scope of research beyond the human discrimination patterns will be important in detecting and evaluating novel AI behaviors. (ii) \textit{Principled AI Approach:} A systematic and principled design approach (with potential use of human oversight) may help tackle the emerging behaviors and scenarios \citep{WEF}.AI may require more diligent and continuous analysis of its actions, mechanisms for intervention and mitigation, as it progresses towards increased autonomy. 

\noindent\textbf{\textit{Beyond Human Biases:}} (i) \textit{Implicitly Ethical AI:} During its implicitly ethical stage, AI discrimination has been shown to possess many characteristics of the human biases (due to factors like historical data use etc).  Thus far, AI fairness techniques have been relying on the human biases as the primary guideline to detect AI bias. As AI becomes increasingly autonomous, this strategy will not be sufficient as novel AI behaviors will emerge.  (ii) \textit{Increased Autonomy:} In the past decade, AI has been fed extensive amounts of data about individuals, which carries bias risk, based on characteristics and patterns that may not even be obvious to humans. AI discrimination may affect the external outcomes through the numerous feedback loops (in some cases foster and self-reward for the original discrimination). Such issues have the potential to reach algorithmic starvation for some groups, which may require novel legal protections \citep{Legit}.

\noindent\textbf{\textit{Detecting Non-Intuitive AI Discrimination:}}  AI relies on large amounts data, complex patterns and feature combinations, in which discrimination is difficult to detect by humans.  In contrast, on many occasions, human biases are easier to discover and even can be visually detected (e.g. gender, age, race etc.). \textit{Correlation Causation Fallacy:} The difficulties of causal inference may lead AI to falsely interpret coincidental associations as causal. Often, not only an individual would not be aware of the discrimination \citep{Network,Wachter}, experts may require specialized detection techniques to identify such cases.  Informal associations, social networks, browsing activity, personality traits (e.g. introversion), images of personal property \cite{MIT_Car}, behavioral patterns have the potential for AI discrimination.  According to FTC, in AI-based recruiting applications, subtle facial expressions, gestures and intonation variations have been used for hiring decisions \citep{forbes,Wachter}. Further, there are concerns of insufficient legal protections for the resulting discrimination \citep{emerging}. This has motivated a wave of new legislation \citep{illinois}. While explainability and transparency principles play an important role in the prevention, the emerging AI discrimination issues will likely become more difficult to detect and mitigate in the near future. 

\noindent\textbf{\textit{Reward \& Feedback Mechanisms:}} Feedback mechanisms  play a key role in the moral functions of the human brain. Deficient emotional feedback and the dysfunction of reward mechanisms derail the moral outcome even when correct ethical judgements can be reached \citep{psychopathy}. Objective functions, reward and feedback mechanisms also play a critical role in controlling the overall AI system behavior and in producing more trustworthy solutions. If not carefully designed at the system-level, they may lead to dangerous outcomes.

\begin{challenge}
Emerging security issues and machine learning attacks pose serious risks. 
\end{challenge}
\noindent Potential security issues are becoming an increasing concern for  highly-critical AI applications in financial services and other industries \citep{DangersAI}. AI models have been shown to have vulnerabilities to machine learning attacks \citep{blackbox,security,Malicious}. Similarly, deep learning applications are known to be sensitive to even small perturbations in the data \citep{blackbox}. Hence, reducing their vulnerability to adversarial attacks and designing robust models have become a major implementation goal \citep{Szegedy,Goodfellow}. Exploring model robustness during possible poisoning, evasion, and inference attacks  is an emerging topic of interest \citep{KDD}. Beyond their security implications, machine learning attacks have the potential to cause serious impact on ethical behavior. According to \citep{darkrobot}, even with minor interventions, a robots behavior can be turned from collaborative to aggressive. So far, there has been limited work on such emerging issues in financial models. 

\begin{challenge}
The growing use of alternative data in financial services AI models is an ethical concern.
\end{challenge}
Lately, alternative and non-traditional data sources have gained significant interest in the financial services industry. Although the initial use was to improve credit scoring for the segments with limited/no credit history,  fintech firms have significantly broadened the scope through the use of data from data brokers and aggregators \citep{Martin}. Alternative data sources include social media, public records, digital footprints, bill payments, transaction data, along with health-indicators, alcohol consumption and other personal data \citep{Experian,FT}. According to the industry reports, data from brokers and aggregators frequently have quality, bias and privacy issues \citep{Martin,Yale}. Furthermore, FTC complaints list numerous unethical practices during the data acquisition, processing and sale \citep{FTC_Data}. According to \citep{FTC_Scam}, some data brokers were charged for selling data for criminal organizations and fraud rings. With the exception of the latest CCPA and CPRA requirements (for data brokers to register with the state), the regulatory oversight over broker data is extremely limited/non-existent, which remains as a serious concern. The use of alternative data sources in AI and machine learning models poses serious threats to ethical AI.

\begin{challenge}
The lack of ethical design practices in the current and next-generation AI applications may be a bigger problem than the long list of technical challenges.
\end{challenge}

\noindent Ethics ties to other critically important topics in AI like causality and rationality \citep{gips,russel}.  According to Pearl, AI's long-term progress will be limited beyond simple analysis without causality \citep{pearl_why}.  Combined with the unpredictability of the emerging autonomous agents, these factors ensure that AI ethics will remain as a serious challenge in the near future.  However, AI is pervasively used in critical functions in countless applications in many industries. The lack of ethical design considerations in the current and next-generation AI applications is a greater concern than the technical challenges.


\section{Conclusions \& Outlook}
\label{sec:conclusion}

\noindent   AI and model development teams face numerous issues in ethical AI solution development, tied to the complexity of the underlying ethics concepts, design complexities, implementation issues, shortage of supporting tools, as well as organizational obstacles. This paper aims to highlight the overarching themes and challenges of building ethical AI solutions in financial services. It aims to bring awareness to the issues and start industry-wide conversations on the potential solution paths. 

\noindent Analyzing the practical challenges considerations enables a number of observations towards actionable steps and solution opportunities: (i) Due to the unpredictability of AI, ethical design practices need to be implemented pervasively across all models and applications.  Current ethics scope for AI deployments is limited and needs to be extended both in terms of the  principles and the application scope. (ii) Defining what is ethical behavior for a broader range of applications requires industry-level discussions and collaboration with regulators and academic experts. (iii) Regulations and novel regulatory frameworks are essential in ensuring and enforcing the desired behavior. Without regulations it is really difficult to reach any ethics goals in the financial services industry. (iv) AI and model development organizations face significant issues because of the gap between high-level guidelines and the AI applications. Firm-wide AI ethics frameworks (similar to risk frameworks), ethics guidelines, processes, standards, tools, methodologies, organizational constructs are needed. (v) AI development teams have limited power in their efforts to develop more ethical solutions. Organizational adjustments and support are essential in ensuring the success of such efforts. (vi) Finally, emerging issues pose serious difficulties in the design of next generation AI applications. They require innovative solutions and collaboration across industry, academia and regulators.

\bibliography{bib}


\begin{thebibliography}{136}


\ifx \showCODEN    \undefined \def \showCODEN     #1{\unskip}     \fi
\ifx \showDOI      \undefined \def \showDOI       #1{#1}\fi
\ifx \showISBNx    \undefined \def \showISBNx     #1{\unskip}     \fi
\ifx \showISBNxiii \undefined \def \showISBNxiii  #1{\unskip}     \fi
\ifx \showISSN     \undefined \def \showISSN      #1{\unskip}     \fi
\ifx \showLCCN     \undefined \def \showLCCN      #1{\unskip}     \fi
\ifx \shownote     \undefined \def \shownote      #1{#1}          \fi
\ifx \showarticletitle \undefined \def \showarticletitle #1{#1}   \fi
\ifx \showURL      \undefined \def \showURL       {\relax}        \fi
\providecommand\bibfield[2]{#2}
\providecommand\bibinfo[2]{#2}
\providecommand\natexlab[1]{#1}
\providecommand\showeprint[2][]{arXiv:#2}

\bibitem[\protect\citeauthoryear{{ACM US Public Policy Council}}{{ACM US Public
  Policy Council}}{2017}]%
        {acm}
\bibfield{author}{\bibinfo{person}{{ACM US Public Policy Council}}.}
  \bibinfo{year}{2017}\natexlab{}.
\newblock \bibinfo{title}{Statement on Algorithmic Transparency and
  Accountability}.
\newblock
\newblock


\bibitem[\protect\citeauthoryear{{AI Now Institute}}{{AI Now
  Institute}}{2019}]%
        {GenderRace}
\bibfield{author}{\bibinfo{person}{{AI Now Institute}}.}
  \bibinfo{year}{2019}\natexlab{}.
\newblock \bibinfo{title}{Discriminating Systems}.
\newblock
\newblock


\bibitem[\protect\citeauthoryear{{Aite} and {Experian}}{{Aite} and
  {Experian}}{2018}]%
        {Experian}
\bibfield{author}{\bibinfo{person}{{Aite}} {and} \bibinfo{person}{{Experian}}.}
  \bibinfo{year}{2018}\natexlab{}.
\newblock \bibinfo{title}{Alternative Data Across the Loan Life Cycle}.
\newblock
\newblock


\bibitem[\protect\citeauthoryear{Allen, Varner, and Zinser}{Allen
  et~al\mbox{.}}{2000}]%
        {cmtt}
\bibfield{author}{\bibinfo{person}{Colin Allen}, \bibinfo{person}{Gary Varner},
  {and} \bibinfo{person}{Jason Zinser}.} \bibinfo{year}{2000}\natexlab{}.
\newblock \showarticletitle{Prolegomena to any future artificial moral agent}.
\newblock \bibinfo{journal}{\emph{J. Exp. Theor. AI}} \bibinfo{volume}{12},
  \bibinfo{number}{3} (\bibinfo{year}{2000}), \bibinfo{pages}{251--61}.
\newblock
\urldef\tempurl%
\url{https://doi.org/10.1080/09528130050111428}
\showDOI{\tempurl}


\bibitem[\protect\citeauthoryear{Anderson and Anderson}{Anderson and
  Anderson}{2007}]%
        {AAAI_Anderson}
\bibfield{author}{\bibinfo{person}{Michael Anderson} {and}
  \bibinfo{person}{Susan~Leigh Anderson}.} \bibinfo{year}{2007}\natexlab{}.
\newblock \showarticletitle{Machine Ethics}.
\newblock \bibinfo{journal}{\emph{AAAI Mag.}} \bibinfo{volume}{28},
  \bibinfo{number}{4} (\bibinfo{year}{2007}), \bibinfo{pages}{15--26}.
\newblock
\urldef\tempurl%
\url{https://doi.org/10.1609/aimag.v28i4.2065}
\showDOI{\tempurl}


\bibitem[\protect\citeauthoryear{Anderson and Anderson}{Anderson and
  Anderson}{2018}]%
        {ethical_turing}
\bibfield{author}{\bibinfo{person}{Michael Anderson} {and}
  \bibinfo{person}{Susan~Leigh Anderson}.} \bibinfo{year}{2018}\natexlab{}.
\newblock \showarticletitle{GenEth}.
\newblock \bibinfo{journal}{\emph{Paladyn}} \bibinfo{volume}{9},
  \bibinfo{number}{1} (\bibinfo{year}{2018}), \bibinfo{pages}{337--57}.
\newblock
\urldef\tempurl%
\url{https://doi.org/10.1515/pjbr-2018-0024}
\showDOI{\tempurl}


\bibitem[\protect\citeauthoryear{Anderson}{Anderson}{2008}]%
        {anderson}
\bibfield{author}{\bibinfo{person}{Susan~Leigh Anderson}.}
  \bibinfo{year}{2008}\natexlab{}.
\newblock \showarticletitle{Asimov's ``three laws of robotics'' and machine
  metaethics}.
\newblock \bibinfo{journal}{\emph{AI Soc.}} \bibinfo{volume}{22},
  \bibinfo{number}{4} (\bibinfo{year}{2008}), \bibinfo{pages}{477--93}.
\newblock
\urldef\tempurl%
\url{https://doi.org/10.1007/s00146-007-0094-5}
\showDOI{\tempurl}


\bibitem[\protect\citeauthoryear{Aran, Such, and Criado}{Aran
  et~al\mbox{.}}{2019}]%
        {ferrer}
\bibfield{author}{\bibinfo{person}{Xavier~Ferrer Aran},
  \bibinfo{person}{Jose~M. Such}, {and} \bibinfo{person}{Natalia Criado}.}
  \bibinfo{year}{2019}\natexlab{}.
\newblock \showarticletitle{Attesting Biases and Discrimination using Language
  Semantics}. In \bibinfo{booktitle}{\emph{AAMAS Workshop on Responsible
  Artificial Intelligence Agents}}. \bibinfo{numpages}{9}~pages.
\newblock
\showeprint[arxiv]{1909.04386}~[cs.AI]


\bibitem[\protect\citeauthoryear{Assefa, Dervovic, Mahfouz, Reddy, and
  Veloso}{Assefa et~al\mbox{.}}{2019}]%
        {JPM}
\bibfield{author}{\bibinfo{person}{Samuel Assefa}, \bibinfo{person}{Danial
  Dervovic}, \bibinfo{person}{Mahmoud Mahfouz}, \bibinfo{person}{Prashant
  Reddy}, {and} \bibinfo{person}{Manuela Veloso}.}
  \bibinfo{year}{2019}\natexlab{}.
\newblock \showarticletitle{Generating Synthetic Data in Finance}. In
  \bibinfo{booktitle}{\emph{NeurIPS Workshop on AI for Financial Services}}.
  \bibinfo{numpages}{10}~pages.
\newblock
\urldef\tempurl%
\url{https://doi.org/10.2139/ssrn.3634235}
\showDOI{\tempurl}


\bibitem[\protect\citeauthoryear{Awad et~al\mbox{.}}{Awad
  et~al\mbox{.}}{2018}]%
        {awad}
\bibfield{author}{\bibinfo{person}{Edmond Awad} {et~al\mbox{.}}}
  \bibinfo{year}{2018}\natexlab{}.
\newblock \showarticletitle{The Moral Machine experiment}.
\newblock \bibinfo{journal}{\emph{Nature}} \bibinfo{volume}{563},
  \bibinfo{number}{7729} (\bibinfo{year}{2018}), \bibinfo{pages}{59--64}.
\newblock
\urldef\tempurl%
\url{https://doi.org/10.1038/s41586-018-0637-6}
\showDOI{\tempurl}


\bibitem[\protect\citeauthoryear{Barnes}{Barnes}{2020}]%
        {forbes}
\bibfield{author}{\bibinfo{person}{Patricia Barnes}.}
  \bibinfo{year}{2020}\natexlab{}.
\newblock \showarticletitle{EPIC Asks FTC To Regulate Use Of AI In
  Pre-Employment Screenings}.
\newblock  (\bibinfo{year}{2020}).
\newblock


\bibitem[\protect\citeauthoryear{Barocas and Selbst}{Barocas and
  Selbst}{2016}]%
        {barocas}
\bibfield{author}{\bibinfo{person}{Solon Barocas} {and}
  \bibinfo{person}{Andrew~D Selbst}.} \bibinfo{year}{2016}\natexlab{}.
\newblock \showarticletitle{Big Data's Disparate Impact}.
\newblock \bibinfo{journal}{\emph{Calif. Law Rev.}}  \bibinfo{volume}{104}
  (\bibinfo{year}{2016}), \bibinfo{pages}{671--732}.
\newblock
\urldef\tempurl%
\url{https://doi.org/10.15779/Z38BG31}
\showDOI{\tempurl}


\bibitem[\protect\citeauthoryear{Barreno, Nelson, Joseph, and Tygar}{Barreno
  et~al\mbox{.}}{2010}]%
        {security}
\bibfield{author}{\bibinfo{person}{Marco Barreno}, \bibinfo{person}{Blaine
  Nelson}, \bibinfo{person}{Anthony~D. Joseph}, {and} \bibinfo{person}{J.~D.
  Tygar}.} \bibinfo{year}{2010}\natexlab{}.
\newblock \showarticletitle{The security of machine learning}.
\newblock \bibinfo{journal}{\emph{Mach. Learn.}} \bibinfo{volume}{81},
  \bibinfo{number}{2} (\bibinfo{year}{2010}), \bibinfo{pages}{121--48}.
\newblock
\urldef\tempurl%
\url{https://doi.org/10.1007/s10994-010-5188-5}
\showDOI{\tempurl}


\bibitem[\protect\citeauthoryear{Bartels et~al\mbox{.}}{Bartels
  et~al\mbox{.}}{2015}]%
        {Bartels}
\bibfield{author}{\bibinfo{person}{Daniel~M. Bartels} {et~al\mbox{.}}}
  \bibinfo{year}{2015}\natexlab{}.
\newblock \showarticletitle{Moral Judgment and Decision Making}.
\newblock In \bibinfo{booktitle}{\emph{Handbook of Judgment and Decision
  Making}}, \bibfield{editor}{\bibinfo{person}{G.~Keren} {and}
  \bibinfo{person}{G.~Wu}} (Eds.). \bibinfo{publisher}{Wiley Blackwell},
  \bibinfo{address}{Chichester, UK}, Chapter~17, \bibinfo{pages}{478--515}.
\newblock


\bibitem[\protect\citeauthoryear{{BCBS SIG}}{{BCBS SIG}}{2019}]%
        {Basel2}
\bibfield{author}{\bibinfo{person}{{BCBS SIG}}.}
  \bibinfo{year}{2019}\natexlab{}.
\newblock \bibinfo{title}{Industry workshop on the governance and oversight of
  artificial intelligence and machine learning in financial services}.
\newblock
\newblock


\bibitem[\protect\citeauthoryear{Bellamy et~al\mbox{.}}{Bellamy
  et~al\mbox{.}}{2018}]%
        {Fairness360}
\bibfield{author}{\bibinfo{person}{Rachel K.~E. Bellamy} {et~al\mbox{.}}}
  \bibinfo{year}{2018}\natexlab{}.
\newblock \bibinfo{title}{AI Fairness 360}.
\newblock
\newblock
\showeprint[arxiv]{1810.01943}


\bibitem[\protect\citeauthoryear{Benedictus}{Benedictus}{2018}]%
        {guardian2}
\bibfield{author}{\bibinfo{person}{Leo Benedictus}.}
  \bibinfo{year}{2018}\natexlab{}.
\newblock \showarticletitle{Look at me}.
\newblock \bibinfo{journal}{\emph{Guardian}} (\bibinfo{year}{2018}).
\newblock


\bibitem[\protect\citeauthoryear{Berk, Heidari, Jabbari, Kearns, and Roth}{Berk
  et~al\mbox{.}}{2021}]%
        {berk}
\bibfield{author}{\bibinfo{person}{Richard Berk}, \bibinfo{person}{Hoda
  Heidari}, \bibinfo{person}{Shahin Jabbari}, \bibinfo{person}{Michael Kearns},
  {and} \bibinfo{person}{Aaron Roth}.} \bibinfo{year}{2021}\natexlab{}.
\newblock \showarticletitle{Fairness in Criminal Justice Risk Assessments}.
\newblock \bibinfo{journal}{\emph{Sociol. Meth. Res.}} \bibinfo{volume}{50},
  \bibinfo{number}{1} (\bibinfo{year}{2021}), \bibinfo{pages}{3--44}.
\newblock
\urldef\tempurl%
\url{https://doi.org/10.1177/0049124118782533}
\showDOI{\tempurl}


\bibitem[\protect\citeauthoryear{Borgohain}{Borgohain}{2018}]%
        {india2}
\bibfield{author}{\bibinfo{person}{Ananya Borgohain}.}
  \bibinfo{year}{2018}\natexlab{}.
\newblock \showarticletitle{Leveraging Tech: How Facebook, SMS and GPS can
  determine if you are a reliable loan applicant}.
\newblock \bibinfo{journal}{\emph{India Times}} (\bibinfo{year}{2018}).
\newblock


\bibitem[\protect\citeauthoryear{Bostrom and Yudkowsky}{Bostrom and
  Yudkowsky}{2011}]%
        {Cambridge}
\bibfield{author}{\bibinfo{person}{Nick Bostrom} {and} \bibinfo{person}{Eliezer
  Yudkowsky}.} \bibinfo{year}{2011}\natexlab{}.
\newblock \showarticletitle{Ethics of AI}.
\newblock In \bibinfo{booktitle}{\emph{Handbook of AI}},
  \bibfield{editor}{\bibinfo{person}{William Ramsey} {and}
  \bibinfo{person}{Keith Frankish}} (Eds.). \bibinfo{publisher}{Cambridge},
  Chapter~15, \bibinfo{pages}{316--34}.
\newblock


\bibitem[\protect\citeauthoryear{boyd, Levy, and Marwick}{boyd
  et~al\mbox{.}}{2014}]%
        {Network}
\bibfield{author}{\bibinfo{person}{danah boyd}, \bibinfo{person}{Karen Levy},
  {and} \bibinfo{person}{Alice Marwick}.} \bibinfo{year}{2014}\natexlab{}.
\newblock \showarticletitle{The Networked Nature of Algorithmic
  Discrimination}.
\newblock In \bibinfo{booktitle}{\emph{Data and Discrimination}},
  \bibfield{editor}{\bibinfo{person}{Seeta~Pe{\~n}a Gangadharan}} (Ed.).
  \bibinfo{publisher}{Open Technology Institute}, Chapter~11,
  \bibinfo{pages}{53--7}.
\newblock


\bibitem[\protect\citeauthoryear{Buckley et~al\mbox{.}}{Buckley
  et~al\mbox{.}}{2019}]%
        {DarkSide}
\bibfield{author}{\bibinfo{person}{Ross~P. Buckley} {et~al\mbox{.}}}
  \bibinfo{year}{2019}\natexlab{}.
\newblock \bibinfo{title}{The Dark Side of Digital Financial Transformation}.
\newblock
\newblock
\urldef\tempurl%
\url{https://doi.org/10.2139/ssrn.3478640}
\showDOI{\tempurl}


\bibitem[\protect\citeauthoryear{Buckley, Zetzsche, Arner, and Tang}{Buckley
  et~al\mbox{.}}{2021}]%
        {FinanceAI}
\bibfield{author}{\bibinfo{person}{Ross~P. Buckley}, \bibinfo{person}{Dirk~A.
  Zetzsche}, \bibinfo{person}{Douglas~W. Arner}, {and} \bibinfo{person}{Brian
  Tang}.} \bibinfo{year}{2021}\natexlab{}.
\newblock \showarticletitle{Regulating AI in Finance}.
\newblock \bibinfo{journal}{\emph{Sydney Law Rev.}}  \bibinfo{volume}{43}
  (\bibinfo{year}{2021}), \bibinfo{pages}{43--81}.
\newblock


\bibitem[\protect\citeauthoryear{Buolamwini and Gebru}{Buolamwini and
  Gebru}{2018}]%
        {MIT}
\bibfield{author}{\bibinfo{person}{Joy Buolamwini} {and}
  \bibinfo{person}{Timnit Gebru}.} \bibinfo{year}{2018}\natexlab{}.
\newblock \showarticletitle{Gender Shades}. In
  \bibinfo{booktitle}{\emph{Proceedings of the 1st Conference on Fairness,
  Accountability and Transparency}}, Vol.~\bibinfo{volume}{81}.
  \bibinfo{publisher}{PMLR}, \bibinfo{pages}{77--91}.
\newblock


\bibitem[\protect\citeauthoryear{{Bureau of Consumer Financial
  Protection}}{{Bureau of Consumer Financial Protection}}{2020}]%
        {CFPB}
\bibfield{author}{\bibinfo{person}{{Bureau of Consumer Financial Protection}}.}
  \bibinfo{year}{2020}\natexlab{}.
\newblock \bibinfo{title}{Fair Lending Report}.
\newblock
\newblock


\bibitem[\protect\citeauthoryear{Calman}{Calman}{2004}]%
        {Calman}
\bibfield{author}{\bibinfo{person}{K.~C. Calman}.}
  \bibinfo{year}{2004}\natexlab{}.
\newblock \showarticletitle{Evolutionary Ethics}.
\newblock \bibinfo{journal}{\emph{J. Med. Ethics}}  \bibinfo{volume}{30}
  (\bibinfo{year}{2004}), \bibinfo{pages}{366--70}.
\newblock
\urldef\tempurl%
\url{https://doi.org/10.1136/jme.2002.003582}
\showDOI{\tempurl}


\bibitem[\protect\citeauthoryear{Cao}{Cao}{2020}]%
        {Australia}
\bibfield{author}{\bibinfo{person}{Longbing Cao}.}
  \bibinfo{year}{2020}\natexlab{}.
\newblock \bibinfo{title}{AI in Finance}.
\newblock
\newblock
\urldef\tempurl%
\url{https://doi.org/10.2139/ssrn.3647625}
\showDOI{\tempurl}


\bibitem[\protect\citeauthoryear{Carrese and Sugarman}{Carrese and
  Sugarman}{2006}]%
        {Bioethics}
\bibfield{author}{\bibinfo{person}{Joseph~A Carrese} {and}
  \bibinfo{person}{Jeremy Sugarman}.} \bibinfo{year}{2006}\natexlab{}.
\newblock \showarticletitle{The Inescapable Relevance of Bioethics for the
  Practicing Clinician}.
\newblock \bibinfo{journal}{\emph{Medical Ethics}} \bibinfo{volume}{130},
  \bibinfo{number}{6} (\bibinfo{year}{2006}), \bibinfo{pages}{P1864--72}.
\newblock
\urldef\tempurl%
\url{https://doi.org/10.1378/chest.130.6.1864}
\showDOI{\tempurl}


\bibitem[\protect\citeauthoryear{Castelfranchi}{Castelfranchi}{2000}]%
        {lying}
\bibfield{author}{\bibinfo{person}{Cristiano Castelfranchi}.}
  \bibinfo{year}{2000}\natexlab{}.
\newblock \showarticletitle{Artificial liars}.
\newblock \bibinfo{journal}{\emph{Ethics Inf. Tech.}} \bibinfo{volume}{2},
  \bibinfo{number}{2} (\bibinfo{year}{2000}), \bibinfo{pages}{113--9}.
\newblock
\urldef\tempurl%
\url{https://doi.org/10.1023/A:1010025403776}
\showDOI{\tempurl}


\bibitem[\protect\citeauthoryear{CFTC}{CFTC}{2019}]%
        {CFTC}
\bibfield{author}{\bibinfo{person}{CFTC}.} \bibinfo{year}{2019}\natexlab{}.
\newblock \bibinfo{title}{A Primer on AI in Financial Markets}.
\newblock
\newblock


\bibitem[\protect\citeauthoryear{Chouldechova}{Chouldechova}{2017}]%
        {Chouldechova}
\bibfield{author}{\bibinfo{person}{Alexandra Chouldechova}.}
  \bibinfo{year}{2017}\natexlab{}.
\newblock \bibinfo{title}{Fair prediction with disparate impact: A study of
  bias in recidivism prediction instruments}.
\newblock
\newblock
\showeprint[arxiv]{1703.00056}~[stat.AP]


\bibitem[\protect\citeauthoryear{Chouldechova and Roth}{Chouldechova and
  Roth}{2020}]%
        {roth}
\bibfield{author}{\bibinfo{person}{Alexandra Chouldechova} {and}
  \bibinfo{person}{Aaron Roth}.} \bibinfo{year}{2020}\natexlab{}.
\newblock \showarticletitle{A Snapshot of the Frontiers of Fairness in Machine
  Learning}.
\newblock \bibinfo{journal}{\emph{Commun. ACM}} \bibinfo{volume}{63},
  \bibinfo{number}{5} (\bibinfo{year}{2020}), \bibinfo{pages}{82---9}.
\newblock
\urldef\tempurl%
\url{https://doi.org/10.1145/3376898}
\showDOI{\tempurl}


\bibitem[\protect\citeauthoryear{Cima, Tonnaer, and Hauser}{Cima
  et~al\mbox{.}}{2010}]%
        {psychopathy}
\bibfield{author}{\bibinfo{person}{Maaike Cima}, \bibinfo{person}{Franca
  Tonnaer}, {and} \bibinfo{person}{Marc~D. Hauser}.}
  \bibinfo{year}{2010}\natexlab{}.
\newblock \showarticletitle{Psychopaths know right from wrong but don't care}.
\newblock \bibinfo{journal}{\emph{Soc. Cogn. Affect. Neur.}}
  \bibinfo{volume}{5}, \bibinfo{number}{1} (\bibinfo{year}{2010}),
  \bibinfo{pages}{59--67}.
\newblock
\urldef\tempurl%
\url{https://doi.org/10.1093/scan/nsp051}
\showDOI{\tempurl}


\bibitem[\protect\citeauthoryear{Corbett-Davies and Goel}{Corbett-Davies and
  Goel}{2018}]%
        {goel}
\bibfield{author}{\bibinfo{person}{Sam Corbett-Davies} {and}
  \bibinfo{person}{Sharad Goel}.} \bibinfo{year}{2018}\natexlab{}.
\newblock \bibinfo{title}{The Measure and Mismeasure of Fairness}.
\newblock
\newblock
\showeprint[arxiv]{1808.00023}~[cs.CY]


\bibitem[\protect\citeauthoryear{Cummings, Gupta, Kimpara, and
  Morgenstern}{Cummings et~al\mbox{.}}{2019}]%
        {privacy}
\bibfield{author}{\bibinfo{person}{Rachel Cummings}, \bibinfo{person}{Varun
  Gupta}, \bibinfo{person}{Dhamma Kimpara}, {and} \bibinfo{person}{Jamie
  Morgenstern}.} \bibinfo{year}{2019}\natexlab{}.
\newblock \showarticletitle{On the Compatibility of Privacy and Fairness}. In
  \bibinfo{booktitle}{\emph{Adjunct Publication of the 27th Conference on User
  Modeling, Adaptation and Personalization}} (Larnaca, Cyprus)
  \emph{(\bibinfo{series}{UMAP'19 Adjunct})}. \bibinfo{pages}{309--15}.
\newblock
\urldef\tempurl%
\url{https://doi.org/10.1145/3314183.3323847}
\showDOI{\tempurl}


\bibitem[\protect\citeauthoryear{D'Amour, Srinivasan, Atwood, Baljekar,
  Sculley, and Halpern}{D'Amour et~al\mbox{.}}{2020}]%
        {google_gym}
\bibfield{author}{\bibinfo{person}{Alexander D'Amour}, \bibinfo{person}{Hansa
  Srinivasan}, \bibinfo{person}{James Atwood}, \bibinfo{person}{Pallavi
  Baljekar}, \bibinfo{person}{D. Sculley}, {and} \bibinfo{person}{Yoni
  Halpern}.} \bibinfo{year}{2020}\natexlab{}.
\newblock \showarticletitle{Fairness is Not Static}. In
  \bibinfo{booktitle}{\emph{Proceedings of the 2020 Conference on Fairness,
  Accountability, and Transparency}} (Barcelona, Spain)
  \emph{(\bibinfo{series}{FAT* '20})}. \bibinfo{pages}{525---34}.
\newblock
\urldef\tempurl%
\url{https://doi.org/10.1145/3351095.3372878}
\showDOI{\tempurl}


\bibitem[\protect\citeauthoryear{Davis, Kumiega, and Van~Vliet}{Davis
  et~al\mbox{.}}{2013}]%
        {HFTEthics}
\bibfield{author}{\bibinfo{person}{Michael Davis}, \bibinfo{person}{Andrew
  Kumiega}, {and} \bibinfo{person}{Ben Van~Vliet}.}
  \bibinfo{year}{2013}\natexlab{}.
\newblock \showarticletitle{Ethics, Finance, and Automation}.
\newblock \bibinfo{journal}{\emph{Sci. Eng. Ethics}} \bibinfo{volume}{19},
  \bibinfo{number}{3} (\bibinfo{year}{2013}), \bibinfo{pages}{851--74}.
\newblock
\urldef\tempurl%
\url{https://doi.org/10.1007/s11948-012-9412-5}
\showDOI{\tempurl}


\bibitem[\protect\citeauthoryear{{Deloitte}}{{Deloitte}}{2019}]%
        {Deloitte}
\bibfield{author}{\bibinfo{person}{{Deloitte}}.}
  \bibinfo{year}{2019}\natexlab{}.
\newblock \bibinfo{title}{Regulating AI in the Banking Space}.
\newblock
\newblock


\bibitem[\protect\citeauthoryear{{Deloitte}}{{Deloitte}}{2020}]%
        {DeloitteDL}
\bibfield{author}{\bibinfo{person}{{Deloitte}}.}
  \bibinfo{year}{2020}\natexlab{}.
\newblock \bibinfo{title}{A Moral License for AI}.
\newblock
\newblock


\bibitem[\protect\citeauthoryear{Deloitte}{Deloitte}{2020}]%
        {Deloitte_AI}
\bibfield{author}{\bibinfo{person}{Deloitte}.} \bibinfo{year}{2020}\natexlab{}.
\newblock \bibinfo{title}{Thriving in the Era of Pervasive AI}.
\newblock
\newblock


\bibitem[\protect\citeauthoryear{Doshi-Velez et~al\mbox{.}}{Doshi-Velez
  et~al\mbox{.}}{2017}]%
        {Accountability}
\bibfield{author}{\bibinfo{person}{Finale Doshi-Velez} {et~al\mbox{.}}}
  \bibinfo{year}{2017}\natexlab{}.
\newblock \bibinfo{title}{Accountability of AI Under the Law}.
\newblock
\newblock
\urldef\tempurl%
\url{https://doi.org/10.2139/ssrn.3064761}
\showDOI{\tempurl}
\showeprint[arxiv]{1711.01134}


\bibitem[\protect\citeauthoryear{Ekstrand, Joshaghani, and Mehrpouyan}{Ekstrand
  et~al\mbox{.}}{2018}]%
        {Ekstrand}
\bibfield{author}{\bibinfo{person}{Michael~D. Ekstrand},
  \bibinfo{person}{Rezvan Joshaghani}, {and} \bibinfo{person}{Hoda
  Mehrpouyan}.} \bibinfo{year}{2018}\natexlab{}.
\newblock \showarticletitle{Privacy for All}. In
  \bibinfo{booktitle}{\emph{Proceedings of the 1st Conference on Fairness,
  Accountability and Transparency}}, Vol.~\bibinfo{volume}{81}.
  \bibinfo{publisher}{PMLR}, \bibinfo{pages}{35--47}.
\newblock


\bibitem[\protect\citeauthoryear{{Ernst and Young}}{{Ernst and Young}}{2019}]%
        {EYFintech}
\bibfield{author}{\bibinfo{person}{{Ernst and Young}}.}
  \bibinfo{year}{2019}\natexlab{}.
\newblock \bibinfo{title}{Global FinTech Adoption Index}.
\newblock
\newblock


\bibitem[\protect\citeauthoryear{{EU AI HLEG}}{{EU AI HLEG}}{2019}]%
        {EUTraceability}
\bibfield{author}{\bibinfo{person}{{EU AI HLEG}}.}
  \bibinfo{year}{2019}\natexlab{}.
\newblock \bibinfo{title}{Ethics Guidelines for Trustworthy AI}.
\newblock
\newblock


\bibitem[\protect\citeauthoryear{Ezrachi and Stucke}{Ezrachi and
  Stucke}{2017}]%
        {Collusion}
\bibfield{author}{\bibinfo{person}{Ariel Ezrachi} {and}
  \bibinfo{person}{Maurice~E. Stucke}.} \bibinfo{year}{2017}\natexlab{}.
\newblock \showarticletitle{AI \& Collusion}.
\newblock \bibinfo{journal}{\emph{U. Ill. Law Rev.}} (\bibinfo{year}{2017}),
  \bibinfo{pages}{1775--1810}.
\newblock


\bibitem[\protect\citeauthoryear{{Federal Trade Commission}}{{Federal Trade
  Commission}}{2015}]%
        {FTC_Scam}
\bibfield{author}{\bibinfo{person}{{Federal Trade Commission}}.}
  \bibinfo{year}{2015}\natexlab{}.
\newblock \bibinfo{title}{FTC Charges Data Brokers with Helping Scammer Take
  More Than \$7 Million from Consumers' Accounts}.
\newblock
\newblock


\bibitem[\protect\citeauthoryear{FINRA}{FINRA}{2020}]%
        {FINRA}
\bibfield{author}{\bibinfo{person}{FINRA}.} \bibinfo{year}{2020}\natexlab{}.
\newblock \bibinfo{title}{AI in the Securities Industry}.
\newblock , \bibinfo{numpages}{20}~pages.
\newblock


\bibitem[\protect\citeauthoryear{Fjeld, Achten, Hilligoss, Nagy, and
  Srikumar}{Fjeld et~al\mbox{.}}{2020}]%
        {harvard}
\bibfield{author}{\bibinfo{person}{Jessica Fjeld}, \bibinfo{person}{Nele
  Achten}, \bibinfo{person}{Hannah Hilligoss}, \bibinfo{person}{Adam Nagy},
  {and} \bibinfo{person}{Madhulika Srikumar}.} \bibinfo{year}{2020}\natexlab{}.
\newblock \bibinfo{booktitle}{\emph{Principled AI}}.
\newblock
\urldef\tempurl%
\url{https://doi.org/10.2139/ssrn.3518482}
\showDOI{\tempurl}


\bibitem[\protect\citeauthoryear{Friedler, Scheidegger, and
  Venkatasubramanian}{Friedler et~al\mbox{.}}{2021}]%
        {impossibilityoffairness}
\bibfield{author}{\bibinfo{person}{Sorelle~A. Friedler},
  \bibinfo{person}{Carlos Scheidegger}, {and} \bibinfo{person}{Suresh
  Venkatasubramanian}.} \bibinfo{year}{2021}\natexlab{}.
\newblock \showarticletitle{The (Im)possibility of Fairness}.
\newblock \bibinfo{journal}{\emph{Commun. ACM}} \bibinfo{volume}{64},
  \bibinfo{number}{4} (\bibinfo{year}{2021}), \bibinfo{pages}{136--43}.
\newblock
\urldef\tempurl%
\url{https://doi.org/10.1145/3433949}
\showDOI{\tempurl}


\bibitem[\protect\citeauthoryear{{Frontier of Humanity Institute}}{{Frontier of
  Humanity Institute}}{2018}]%
        {Malicious}
\bibfield{author}{\bibinfo{person}{{Frontier of Humanity Institute}}.}
  \bibinfo{year}{2018}\natexlab{}.
\newblock \bibinfo{title}{Malicious AI Use Forecasting, Prevention,
  Mitigation}.
\newblock
\newblock


\bibitem[\protect\citeauthoryear{{Future of Life Institute}}{{Future of Life
  Institute}}{2017}]%
        {asilomar}
\bibfield{author}{\bibinfo{person}{{Future of Life Institute}}.}
  \bibinfo{year}{2017}\natexlab{}.
\newblock \bibinfo{title}{Asilomar AI Principles}.
\newblock
\newblock


\bibitem[\protect\citeauthoryear{Gips}{Gips}{1995}]%
        {gips}
\bibfield{author}{\bibinfo{person}{James Gips}.}
  \bibinfo{year}{1995}\natexlab{}.
\newblock \showarticletitle{Towards the Ethical Robot}.
\newblock In \bibinfo{booktitle}{\emph{Android Epistemology}},
  \bibfield{editor}{\bibinfo{person}{Kenneth~M. Ford}, \bibinfo{person}{Clark
  Glymour}, {and} \bibinfo{person}{Patrick Hayes}} (Eds.).
  \bibinfo{publisher}{AAAI}, Chapter~12, \bibinfo{pages}{243--52}.
\newblock


\bibitem[\protect\citeauthoryear{Goodfellow, Shlens, and Szegedy}{Goodfellow
  et~al\mbox{.}}{2015}]%
        {Goodfellow}
\bibfield{author}{\bibinfo{person}{Ian Goodfellow}, \bibinfo{person}{Jonathon
  Shlens}, {and} \bibinfo{person}{Christian Szegedy}.}
  \bibinfo{year}{2015}\natexlab{}.
\newblock \showarticletitle{Explaining \& Harnessing Adversarial Examples}. In
  \bibinfo{booktitle}{\emph{ICLR}}.
\newblock


\bibitem[\protect\citeauthoryear{Grover}{Grover}{5  6}]%
        {Legit}
\bibfield{author}{\bibinfo{person}{Susan Grover}.}
  \bibinfo{year}{1995--6}\natexlab{}.
\newblock \showarticletitle{The Business Necessity Defense in Disparate Impact
  Discrimination Cases}.
\newblock \bibinfo{journal}{\emph{Georgia Law Rev.}}  \bibinfo{volume}{30}
  (\bibinfo{year}{1995--6}), \bibinfo{pages}{387--430}.
\newblock


\bibitem[\protect\citeauthoryear{Gulamhuseinwala et~al\mbox{.}}{Gulamhuseinwala
  et~al\mbox{.}}{15}]%
        {Gulam}
\bibfield{author}{\bibinfo{person}{Imran Gulamhuseinwala} {et~al\mbox{.}}}
  \bibinfo{year}{15}\natexlab{}.
\newblock \showarticletitle{FinTech is Gaining Traction and Young, High-Income
  Users are the Early Adopters}.
\newblock \bibinfo{journal}{\emph{J. Finan. Perspect.}} \bibinfo{volume}{3},
  \bibinfo{number}{3} (\bibinfo{year}{15}), \bibinfo{pages}{16--23}.
\newblock


\bibitem[\protect\citeauthoryear{Gunning and Aha}{Gunning and Aha}{2019}]%
        {DARPA}
\bibfield{author}{\bibinfo{person}{David Gunning} {and} \bibinfo{person}{David
  Aha}.} \bibinfo{year}{2019}\natexlab{}.
\newblock \showarticletitle{DARPA's Explainable Artificial Intelligence (XAI)
  Program}.
\newblock \bibinfo{journal}{\emph{AI Mag.}} \bibinfo{volume}{40},
  \bibinfo{number}{2} (\bibinfo{year}{2019}), \bibinfo{pages}{44--58}.
\newblock
\urldef\tempurl%
\url{https://doi.org/10.1609/aimag.v40i2.2850}
\showDOI{\tempurl}


\bibitem[\protect\citeauthoryear{Hagendorff}{Hagendorff}{2020}]%
        {EthicsofAIEthics}
\bibfield{author}{\bibinfo{person}{Thilo Hagendorff}.}
  \bibinfo{year}{2020}\natexlab{}.
\newblock \showarticletitle{The Ethics of AI Ethics}.
\newblock \bibinfo{journal}{\emph{Minds Mach.}} \bibinfo{volume}{30},
  \bibinfo{number}{1} (\bibinfo{year}{2020}), \bibinfo{pages}{99--120}.
\newblock
\urldef\tempurl%
\url{https://doi.org/10.1007/s11023-020-09517-8}
\showDOI{\tempurl}


\bibitem[\protect\citeauthoryear{Hagerty and Rubinov}{Hagerty and
  Rubinov}{2019}]%
        {AIImpact}
\bibfield{author}{\bibinfo{person}{Alexa Hagerty} {and} \bibinfo{person}{Igor
  Rubinov}.} \bibinfo{year}{2019}\natexlab{}.
\newblock \bibinfo{title}{Global AI Ethics}.
\newblock
\newblock
\showeprint[arxiv]{1907.07892}~[cs.CY]


\bibitem[\protect\citeauthoryear{Hao}{Hao}{2019}]%
        {MIT_bias}
\bibfield{author}{\bibinfo{person}{Karen Hao}.}
  \bibinfo{year}{2019}\natexlab{}.
\newblock \showarticletitle{This is How AI Bias Really Happens}.
\newblock \bibinfo{journal}{\emph{MIT Tech. Rev.}} (\bibinfo{year}{2019}).
\newblock


\bibitem[\protect\citeauthoryear{Holland, Hosny, Newman, Joseph, and
  Chmielinski}{Holland et~al\mbox{.}}{2018}]%
        {DataQuality}
\bibfield{author}{\bibinfo{person}{Sarah Holland}, \bibinfo{person}{Ahmed
  Hosny}, \bibinfo{person}{Sarah Newman}, \bibinfo{person}{Joshua Joseph},
  {and} \bibinfo{person}{Kasia Chmielinski}.} \bibinfo{year}{2018}\natexlab{}.
\newblock \showarticletitle{The Dataset Nutrition Label}.
\newblock
\showeprint[arxiv]{1805.03677}~[cs.DB]


\bibitem[\protect\citeauthoryear{Hooker, Moorosi, Clark, Bengio, and
  Denton}{Hooker et~al\mbox{.}}{2020}]%
        {hooker2020characterising}
\bibfield{author}{\bibinfo{person}{Sara Hooker}, \bibinfo{person}{Nyalleng
  Moorosi}, \bibinfo{person}{Gregory Clark}, \bibinfo{person}{Samy Bengio},
  {and} \bibinfo{person}{Emily Denton}.} \bibinfo{year}{2020}\natexlab{}.
\newblock \bibinfo{title}{Characterising Bias in Compressed Models}.
\newblock
\newblock
\showeprint[arxiv]{2010.03058}


\bibitem[\protect\citeauthoryear{Hu, Immorlica, and Vaughan}{Hu
  et~al\mbox{.}}{2019}]%
        {Hu}
\bibfield{author}{\bibinfo{person}{Lily Hu}, \bibinfo{person}{Nicole
  Immorlica}, {and} \bibinfo{person}{Jennifer~Wortman Vaughan}.}
  \bibinfo{year}{2019}\natexlab{}.
\newblock \showarticletitle{The Disparate Effects of Strategic Manipulation}.
  In \bibinfo{booktitle}{\emph{Proceedings of the Conference on Fairness,
  Accountability, and Transparency}} (Atlanta, GA) \emph{(\bibinfo{series}{FAT*
  '19})}. \bibinfo{pages}{259--268}.
\newblock
\urldef\tempurl%
\url{https://doi.org/10.1145/3287560.3287597}
\showDOI{\tempurl}


\bibitem[\protect\citeauthoryear{Hunt}{Hunt}{2016}]%
        {microsoft2}
\bibfield{author}{\bibinfo{person}{Elle Hunt}.}
  \bibinfo{year}{2016}\natexlab{}.
\newblock \showarticletitle{Tay, Microsoft's AI chatbot, gets a crash course in
  racism from Twitter}.
\newblock \bibinfo{journal}{\emph{Guardian}} (\bibinfo{year}{2016}).
\newblock


\bibitem[\protect\citeauthoryear{{IEEE Global Initiative on Ethics of
  Autonomous and Intelligent Systems}}{{IEEE Global Initiative on Ethics of
  Autonomous and Intelligent Systems}}{2017}]%
        {IEEE_Principles}
\bibfield{author}{\bibinfo{person}{{IEEE Global Initiative on Ethics of
  Autonomous and Intelligent Systems}}.} \bibinfo{year}{2017}\natexlab{}.
\newblock \showarticletitle{Ethically Aligned Design --- Version {II}}.
\newblock  (\bibinfo{year}{2017}).
\newblock


\bibitem[\protect\citeauthoryear{{Illinois General Assembly}}{{Illinois General
  Assembly}}{2020}]%
        {illinois}
\bibfield{author}{\bibinfo{person}{{Illinois General Assembly}}.}
  \bibinfo{year}{2020}\natexlab{}.
\newblock \bibinfo{title}{Public Act 101-0260: Artificial Intelligence Video
  Interview Act}.
\newblock
\newblock


\bibitem[\protect\citeauthoryear{Jobin, Ienca, and Vayena}{Jobin
  et~al\mbox{.}}{2019}]%
        {ETH}
\bibfield{author}{\bibinfo{person}{Anna Jobin}, \bibinfo{person}{Marcello
  Ienca}, {and} \bibinfo{person}{Effy Vayena}.}
  \bibinfo{year}{2019}\natexlab{}.
\newblock \showarticletitle{The global landscape of AI ethics guidelines}.
\newblock \bibinfo{journal}{\emph{Nat. Mach. Intell.}} \bibinfo{volume}{1},
  \bibinfo{number}{9} (\bibinfo{year}{2019}), \bibinfo{pages}{389--99}.
\newblock
\urldef\tempurl%
\url{https://doi.org/10.1038/s42256-019-0088-2}
\showDOI{\tempurl}


\bibitem[\protect\citeauthoryear{Kahane}{Kahane}{2015}]%
        {Kahane1}
\bibfield{author}{\bibinfo{person}{Guy Kahane}.}
  \bibinfo{year}{2015}\natexlab{}.
\newblock \showarticletitle{Sidetracked by trolleys}.
\newblock \bibinfo{journal}{\emph{Soc. Neurosci.}} \bibinfo{volume}{10},
  \bibinfo{number}{5} (\bibinfo{year}{2015}), \bibinfo{pages}{551--60}.
\newblock
\urldef\tempurl%
\url{https://doi.org/10.1080/17470919.2015.1023400}
\showDOI{\tempurl}


\bibitem[\protect\citeauthoryear{Karppi and Crawford}{Karppi and
  Crawford}{2016}]%
        {Karppi}
\bibfield{author}{\bibinfo{person}{Tero Karppi} {and} \bibinfo{person}{Kate
  Crawford}.} \bibinfo{year}{2016}\natexlab{}.
\newblock \showarticletitle{Social Media, Financial Algorithms and the Hack
  Crash}.
\newblock \bibinfo{journal}{\emph{Theory Culture Soc.}} \bibinfo{volume}{33},
  \bibinfo{number}{1} (\bibinfo{year}{2016}), \bibinfo{pages}{73--92}.
\newblock
\urldef\tempurl%
\url{https://doi.org/10.1177/0263276415583139}
\showDOI{\tempurl}


\bibitem[\protect\citeauthoryear{Kaur, Nori, Jenkins, Caruana, Wallach, and
  Wortman~Vaughan}{Kaur et~al\mbox{.}}{2020}]%
        {10.1145/3313831.3376219}
\bibfield{author}{\bibinfo{person}{Harmanpreet Kaur}, \bibinfo{person}{Harsha
  Nori}, \bibinfo{person}{Samuel Jenkins}, \bibinfo{person}{Rich Caruana},
  \bibinfo{person}{Hanna Wallach}, {and} \bibinfo{person}{Jennifer
  Wortman~Vaughan}.} \bibinfo{year}{2020}\natexlab{}.
\newblock \showarticletitle{Interpreting Interpretability}. In
  \bibinfo{booktitle}{\emph{Proceedings of the 2020 CHI Conference on Human
  Factors in Computing Systems}} (Honolulu, HI) \emph{(\bibinfo{series}{CHI
  '20})}. \bibinfo{pages}{1--14}.
\newblock
\urldef\tempurl%
\url{https://doi.org/10.1145/3313831.3376219}
\showDOI{\tempurl}


\bibitem[\protect\citeauthoryear{Kita and Kidzi{\'n}ski}{Kita and
  Kidzi{\'n}ski}{2019}]%
        {MIT_Car}
\bibfield{author}{\bibinfo{person}{Kinga Kita} {and} \bibinfo{person}{{\L}ukasz
  Kidzi{\'n}ski}.} \bibinfo{year}{2019}\natexlab{}.
\newblock \bibinfo{title}{Google Street View image of a house predicts car
  accident risk of its resident}.
\newblock
\newblock
\showeprint[arxiv]{1904.05270}~[stat.AP]


\bibitem[\protect\citeauthoryear{Kleinberg, Mullainathan, and
  Raghavan}{Kleinberg et~al\mbox{.}}{2017}]%
        {kleinberg}
\bibfield{author}{\bibinfo{person}{Jon Kleinberg}, \bibinfo{person}{Sendhil
  Mullainathan}, {and} \bibinfo{person}{Manish Raghavan}.}
  \bibinfo{year}{2017}\natexlab{}.
\newblock \showarticletitle{Inherent Trade-Offs in the Fair Determination of
  Risk Scores}. In \bibinfo{booktitle}{\emph{8th Innovations in Theoretical
  Computer Science Conference (ITCS 2017)}},
  \bibfield{editor}{\bibinfo{person}{Christos~H. Papadimitriou}} (Ed.),
  Vol.~\bibinfo{volume}{67}. \bibinfo{publisher}{Schloss Dagstuhl},
  \bibinfo{address}{Dagstuhl, Germany}, \bibinfo{pages}{43:1--23}.
\newblock
\urldef\tempurl%
\url{https://doi.org/10.4230/LIPIcs.ITCS.2017.43}
\showDOI{\tempurl}


\bibitem[\protect\citeauthoryear{Knight}{Knight}{2019}]%
        {applecard}
\bibfield{author}{\bibinfo{person}{Will Knight}.}
  \bibinfo{year}{2019}\natexlab{}.
\newblock \showarticletitle{The Apple Card Didn't `See' Gender}.
\newblock \bibinfo{journal}{\emph{Wired}} (\bibinfo{year}{2019}).
\newblock


\bibitem[\protect\citeauthoryear{{KPMG}}{{KPMG}}{2018}]%
        {kpmg}
\bibfield{author}{\bibinfo{person}{{KPMG}}.} \bibinfo{year}{2018}\natexlab{}.
\newblock \bibinfo{title}{Forging the Future}.
\newblock
\newblock


\bibitem[\protect\citeauthoryear{Kurshan, Shen, and Chen}{Kurshan
  et~al\mbox{.}}{2020}]%
        {ICAIF}
\bibfield{author}{\bibinfo{person}{Eren Kurshan}, \bibinfo{person}{Hongda
  Shen}, {and} \bibinfo{person}{Jiahao Chen}.} \bibinfo{year}{2020}\natexlab{}.
\newblock \showarticletitle{Towards Self-Regulating AI}. In
  \bibinfo{booktitle}{\emph{Proceedings of the First ACM International
  Conference on AI in Finance}} \emph{(\bibinfo{series}{ICAIF '20})}. Article
  \bibinfo{articleno}{49}, \bibinfo{numpages}{8}~pages.
\newblock
\urldef\tempurl%
\url{https://doi.org/10.1145/3383455.3422564}
\showDOI{\tempurl}


\bibitem[\protect\citeauthoryear{Kusner, Loftus, Russell, and Silva}{Kusner
  et~al\mbox{.}}{2017}]%
        {kusner}
\bibfield{author}{\bibinfo{person}{Matt~J Kusner}, \bibinfo{person}{Joshua
  Loftus}, \bibinfo{person}{Chris Russell}, {and} \bibinfo{person}{Ricardo
  Silva}.} \bibinfo{year}{2017}\natexlab{}.
\newblock \showarticletitle{Counterfactual Fairness}. In
  \bibinfo{booktitle}{\emph{Advances in Neural Information Processing
  Systems}}, Vol.~\bibinfo{volume}{30}. \bibinfo{pages}{4066--76}.
\newblock


\bibitem[\protect\citeauthoryear{Leibo, Zambaldi, Lanctot, Marecki, and
  Graepel}{Leibo et~al\mbox{.}}{2017}]%
        {aggression}
\bibfield{author}{\bibinfo{person}{Joel~Z. Leibo}, \bibinfo{person}{Vinicius
  Zambaldi}, \bibinfo{person}{Marc Lanctot}, \bibinfo{person}{Janusz Marecki},
  {and} \bibinfo{person}{Thore Graepel}.} \bibinfo{year}{2017}\natexlab{}.
\newblock \showarticletitle{Multi-Agent Reinforcement Learning in Sequential
  Social Dilemmas}. In \bibinfo{booktitle}{\emph{Proceedings of the 16th
  Conference on Autonomous Agents and MultiAgent Systems}} (S\~{a}o Paulo,
  Brazil) \emph{(\bibinfo{series}{AAMAS '17})}. \bibinfo{address}{Richland,
  SC}, \bibinfo{pages}{464--473}.
\newblock


\bibitem[\protect\citeauthoryear{Lieber}{Lieber}{2009}]%
        {amex}
\bibfield{author}{\bibinfo{person}{Ron Lieber}.}
  \bibinfo{year}{2009}\natexlab{}.
\newblock \showarticletitle{American Express Kept a (Very) Watchful Eye on
  Charges}.
\newblock \bibinfo{journal}{\emph{New York Times}} (\bibinfo{year}{2009}).
\newblock


\bibitem[\protect\citeauthoryear{London}{London}{2019}]%
        {ExplainabilityvsAccuracy}
\bibfield{author}{\bibinfo{person}{Alex~John London}.}
  \bibinfo{year}{2019}\natexlab{}.
\newblock \showarticletitle{AI and Black-Box Medical Decisions}.
\newblock \bibinfo{journal}{\emph{Hastings Center Rep.}} \bibinfo{volume}{49},
  \bibinfo{number}{1} (\bibinfo{year}{2019}), \bibinfo{pages}{15--21}.
\newblock
\urldef\tempurl%
\url{https://doi.org/10.1002/hast.973}
\showDOI{\tempurl}


\bibitem[\protect\citeauthoryear{Madras, Creager, Pitassi, and Zemel}{Madras
  et~al\mbox{.}}{2018}]%
        {madras}
\bibfield{author}{\bibinfo{person}{David Madras}, \bibinfo{person}{Elliot
  Creager}, \bibinfo{person}{Toniann Pitassi}, {and} \bibinfo{person}{Richard
  Zemel}.} \bibinfo{year}{2018}\natexlab{}.
\newblock \showarticletitle{Learning Adversarially Fair and Transferable
  Representations}. In \bibinfo{booktitle}{\emph{Proceedings of the 35th
  International Conference on Machine Learning}}, Vol.~\bibinfo{volume}{80}.
  \bibinfo{publisher}{PMLR}, \bibinfo{pages}{3384--93}.
\newblock


\bibitem[\protect\citeauthoryear{Mann and Matzner}{Mann and Matzner}{2019}]%
        {emerging}
\bibfield{author}{\bibinfo{person}{Monique Mann} {and} \bibinfo{person}{Tobias
  Matzner}.} \bibinfo{year}{2019}\natexlab{}.
\newblock \showarticletitle{Challenging algorithmic profiling}.
\newblock \bibinfo{journal}{\emph{Big Data Soc.}} \bibinfo{volume}{6},
  \bibinfo{number}{2} (\bibinfo{year}{2019}), \bibinfo{numpages}{11}~pages.
\newblock
\urldef\tempurl%
\url{https://doi.org/10.1177/2053951719895805}
\showDOI{\tempurl}


\bibitem[\protect\citeauthoryear{Martin}{Martin}{2015}]%
        {Martin}
\bibfield{author}{\bibinfo{person}{Kirsten~E. Martin}.}
  \bibinfo{year}{2015}\natexlab{}.
\newblock \showarticletitle{Ethical Issues in the Big Data Industry}.
\newblock \bibinfo{journal}{\emph{MIS Q. Exec.}} \bibinfo{volume}{14},
  \bibinfo{number}{2} (\bibinfo{year}{2015}), \bibinfo{numpages}{19}~pages.
\newblock


\bibitem[\protect\citeauthoryear{{McKinsey}}{{McKinsey}}{2017}]%
        {mckinsey_modelcount}
\bibfield{author}{\bibinfo{person}{{McKinsey}}.}
  \bibinfo{year}{2017}\natexlab{}.
\newblock \bibinfo{title}{The Evolution of Model Risk Management}.
\newblock
\newblock


\bibitem[\protect\citeauthoryear{Mehrabi et~al\mbox{.}}{Mehrabi
  et~al\mbox{.}}{2021}]%
        {mehrabi}
\bibfield{author}{\bibinfo{person}{Ninareh Mehrabi} {et~al\mbox{.}}}
  \bibinfo{year}{2021}\natexlab{}.
\newblock \showarticletitle{A Survey on Bias and Fairness in Machine Learning}.
\newblock \bibinfo{journal}{\emph{ACM Comput. Surv.}} \bibinfo{volume}{54},
  \bibinfo{number}{6}, Article \bibinfo{articleno}{115} (\bibinfo{date}{July}
  \bibinfo{year}{2021}), \bibinfo{numpages}{35}~pages.
\newblock
\urldef\tempurl%
\url{https://doi.org/10.1145/3457607}
\showDOI{\tempurl}


\bibitem[\protect\citeauthoryear{Milli, Miller, Dragan, and Hardt}{Milli
  et~al\mbox{.}}{2019}]%
        {hardt18}
\bibfield{author}{\bibinfo{person}{Smitha Milli}, \bibinfo{person}{John
  Miller}, \bibinfo{person}{Anca~D. Dragan}, {and} \bibinfo{person}{Moritz
  Hardt}.} \bibinfo{year}{2019}\natexlab{}.
\newblock \showarticletitle{The Social Cost of Strategic Classification}. In
  \bibinfo{booktitle}{\emph{Proceedings of the Conference on Fairness,
  Accountability, and Transparency}} (Atlanta, GA) \emph{(\bibinfo{series}{FAT*
  '19})}. \bibinfo{pages}{230---9}.
\newblock
\urldef\tempurl%
\url{https://doi.org/10.1145/3287560.3287576}
\showDOI{\tempurl}


\bibitem[\protect\citeauthoryear{Mittelstadt}{Mittelstadt}{2019}]%
        {Principles}
\bibfield{author}{\bibinfo{person}{Brent Mittelstadt}.}
  \bibinfo{year}{2019}\natexlab{}.
\newblock \showarticletitle{Principles Alone Can't Guarantee Ethical AI}.
\newblock \bibinfo{journal}{\emph{Nat. Mach. Intell.}} (\bibinfo{year}{2019}).
\newblock
\urldef\tempurl%
\url{https://doi.org/10.1038/s42256-019-0114-4}
\showDOI{\tempurl}


\bibitem[\protect\citeauthoryear{Mnih et~al\mbox{.}}{Mnih
  et~al\mbox{.}}{2015}]%
        {MnihKSR2015}
\bibfield{author}{\bibinfo{person}{Volodymyr Mnih} {et~al\mbox{.}}}
  \bibinfo{year}{2015}\natexlab{}.
\newblock \showarticletitle{Human-level control through deep reinforcement
  learning}.
\newblock \bibinfo{journal}{\emph{Nature}} \bibinfo{volume}{518},
  \bibinfo{number}{7540} (\bibinfo{year}{2015}), \bibinfo{pages}{529--33}.
\newblock
\urldef\tempurl%
\url{https://doi.org/10.1038/nature14236}
\showDOI{\tempurl}


\bibitem[\protect\citeauthoryear{M{\"u}ller}{M{\"u}ller}{2020}]%
        {Stanford_AI}
\bibfield{author}{\bibinfo{person}{Vincent~C. M{\"u}ller}.}
  \bibinfo{year}{2020}\natexlab{}.
\newblock \showarticletitle{Ethics of AI and Robotics}.
\newblock In \bibinfo{booktitle}{\emph{Encycl. Phil.}},
  \bibfield{editor}{\bibinfo{person}{Edward~N. Zalta}} (Ed.).
  \bibinfo{publisher}{Stanford}.
\newblock


\bibitem[\protect\citeauthoryear{Murgia and Shrikanth}{Murgia and
  Shrikanth}{2019}]%
        {Financialtimes}
\bibfield{author}{\bibinfo{person}{Madhumita Murgia} {and}
  \bibinfo{person}{Siddarth Shrikanth}.} \bibinfo{year}{2019}\natexlab{}.
\newblock \showarticletitle{How Big Tech is struggling with the ethics of AI}.
\newblock \bibinfo{journal}{\emph{Finan. Times}} (\bibinfo{year}{2019}).
\newblock


\bibitem[\protect\citeauthoryear{Narayanan}{Narayanan}{2018}]%
        {Narayanan}
\bibfield{author}{\bibinfo{person}{Arvind Narayanan}.}
  \bibinfo{year}{2018}\natexlab{}.
\newblock \showarticletitle{21 Fairness Definitions and Their Politics}.
\newblock \bibinfo{journal}{\emph{FAT*}} (\bibinfo{year}{2018}).
\newblock


\bibitem[\protect\citeauthoryear{{\'O}Broin and O'Riordan}{{\'O}Broin and
  O'Riordan}{2007}]%
        {deception}
\bibfield{author}{\bibinfo{person}{Pilib {\'O}Broin} {and}
  \bibinfo{person}{Colm O'Riordan}.} \bibinfo{year}{2007}\natexlab{}.
\newblock \showarticletitle{An evolutionary approach to deception in
  multi-agent systems}.
\newblock \bibinfo{journal}{\emph{AI Rev.}} \bibinfo{volume}{27},
  \bibinfo{number}{4} (\bibinfo{year}{2007}), \bibinfo{pages}{257--71}.
\newblock
\urldef\tempurl%
\url{https://doi.org/10.1007/s10462-008-9080-7}
\showDOI{\tempurl}


\bibitem[\protect\citeauthoryear{{OCC Bulletin 2020-59}}{{OCC Bulletin
  2020-59}}{2020}]%
        {OCC}
\bibfield{author}{\bibinfo{person}{{OCC Bulletin 2020-59}}.}
  \bibinfo{year}{2020}\natexlab{}.
\newblock \bibinfo{title}{National Bank and Federal Savings Association Digital
  Activities: Advance Notice of Proposed Rulemaking}.
\newblock
\newblock


\bibitem[\protect\citeauthoryear{OECD Report}{OECD Report}{2019}]%
        {OECD}
OECD Report \bibinfo{year}{2019}\natexlab{}.
\newblock \bibinfo{title}{OECD Principles on AI}.
\newblock
\newblock


\bibitem[\protect\citeauthoryear{Papernot et~al\mbox{.}}{Papernot
  et~al\mbox{.}}{2017}]%
        {blackbox}
\bibfield{author}{\bibinfo{person}{Nicolas Papernot} {et~al\mbox{.}}}
  \bibinfo{year}{2017}\natexlab{}.
\newblock \showarticletitle{Practical Black-Box Attacks against Machine
  Learning}. In \bibinfo{booktitle}{\emph{Proceedings of the 2017 ACM Asia
  Conference on Computer and Communications Security}} (Abu Dhabi, UAE)
  \emph{(\bibinfo{series}{ASIA CCS '17})}. \bibinfo{pages}{506--19}.
\newblock
\urldef\tempurl%
\url{https://doi.org/10.1145/3052973.3053009}
\showDOI{\tempurl}


\bibitem[\protect\citeauthoryear{Passi and Barocas}{Passi and Barocas}{2019}]%
        {Solon}
\bibfield{author}{\bibinfo{person}{Samir Passi} {and} \bibinfo{person}{Solon
  Barocas}.} \bibinfo{year}{2019}\natexlab{}.
\newblock \showarticletitle{Problem Formulation and Fairness}. In
  \bibinfo{booktitle}{\emph{Proceedings of the Conference on Fairness,
  Accountability, and Transparency}} (Atlanta, GA) \emph{(\bibinfo{series}{FAT*
  '19})}. \bibinfo{pages}{39--48}.
\newblock
\showISBNx{9781450361255}
\urldef\tempurl%
\url{https://doi.org/10.1145/3287560.3287567}
\showDOI{\tempurl}


\bibitem[\protect\citeauthoryear{Pearl}{Pearl}{2018}]%
        {pearl_why}
\bibfield{author}{\bibinfo{person}{Judea Pearl}.}
  \bibinfo{year}{2018}\natexlab{}.
\newblock \bibinfo{booktitle}{\emph{The Book of Why}}.
\newblock \bibinfo{publisher}{Basic Books}.
\newblock


\bibitem[\protect\citeauthoryear{Prabhakar}{Prabhakar}{2020}]%
        {berkeley2}
\bibfield{author}{\bibinfo{person}{Tarunima Prabhakar}.}
  \bibinfo{year}{2020}\natexlab{}.
\newblock \bibinfo{booktitle}{\emph{A new era for credit scoring}}.
\newblock \bibinfo{publisher}{Berkeley Center for Long-Term Cybersecurity}.
\newblock


\bibitem[\protect\citeauthoryear{Prinz}{Prinz}{2009}]%
        {EmotionalMoral}
\bibfield{author}{\bibinfo{person}{Jesse Prinz}.}
  \bibinfo{year}{2009}\natexlab{}.
\newblock \bibinfo{booktitle}{\emph{The Emotional Construction of Morals}}.
\newblock \bibinfo{publisher}{Oxford Scholarship}.
\newblock
\urldef\tempurl%
\url{https://doi.org/10.1093/acprof:oso/9780199571543.001.0001}
\showDOI{\tempurl}


\bibitem[\protect\citeauthoryear{Ram and Murgia}{Ram and Murgia}{2019}]%
        {FT}
\bibfield{author}{\bibinfo{person}{A. Ram} {and} \bibinfo{person}{M. Murgia}.}
  \bibinfo{year}{2019}\natexlab{}.
\newblock \showarticletitle{Data Brokers}.
\newblock \bibinfo{journal}{\emph{Finan. Times}} (\bibinfo{year}{2019}).
\newblock


\bibitem[\protect\citeauthoryear{Ramirez, Brill, Ohlhausen, and
  McSweeny}{Ramirez et~al\mbox{.}}{2016}]%
        {FTC}
\bibfield{author}{\bibinfo{person}{Edith Ramirez}, \bibinfo{person}{Julie
  Brill}, \bibinfo{person}{Maureen~K. Ohlhausen}, {and}
  \bibinfo{person}{Terrell McSweeny}.} \bibinfo{year}{2016}\natexlab{}.
\newblock \bibinfo{booktitle}{\emph{Big Data: A Tool for Inclusion or
  Exclusion?}}
\newblock \bibinfo{publisher}{Federal Trade Commission}.
\newblock


\bibitem[\protect\citeauthoryear{Ramirez, Brill, Ohlhausen, Wright, and
  McSweeny}{Ramirez et~al\mbox{.}}{2014}]%
        {FTC_Data}
\bibfield{author}{\bibinfo{person}{Edith Ramirez}, \bibinfo{person}{Julie
  Brill}, \bibinfo{person}{Maureen~K. Ohlhausen}, \bibinfo{person}{Joshua~D.
  Wright}, {and} \bibinfo{person}{Terrell McSweeny}.}
  \bibinfo{year}{2014}\natexlab{}.
\newblock \bibinfo{booktitle}{\emph{Data Brokers: A Call for Transparency and
  Accountability}}.
\newblock \bibinfo{publisher}{Federal Trade Commission}.
\newblock


\bibitem[\protect\citeauthoryear{Rea}{Rea}{2020}]%
        {capone}
\bibfield{author}{\bibinfo{person}{Stephen Rea}.}
  \bibinfo{year}{2020}\natexlab{}.
\newblock \bibinfo{title}{A Survey of Fair and Responsible ML and AI:
  Implications of Consumer Financial Services}.
\newblock , \bibinfo{numpages}{39}~pages.
\newblock
\urldef\tempurl%
\url{https://doi.org/10.2139/ssrn.3527034}
\showDOI{\tempurl}


\bibitem[\protect\citeauthoryear{Rebello and Shukla}{Rebello and
  Shukla}{2017}]%
        {india1}
\bibfield{author}{\bibinfo{person}{Joel Rebello} {and} \bibinfo{person}{Saloni
  Shukla}.} \bibinfo{year}{2017}\natexlab{}.
\newblock \showarticletitle{Forget credit rating, your social media posts may
  decide whether you will get a loan or not}.
\newblock \bibinfo{journal}{\emph{Econ. Times (India)}} (\bibinfo{year}{2017}).
\newblock


\bibitem[\protect\citeauthoryear{Rostow}{Rostow}{2017}]%
        {Yale}
\bibfield{author}{\bibinfo{person}{Theodore Rostow}.}
  \bibinfo{year}{2017}\natexlab{}.
\newblock \showarticletitle{What Happens When an Acquaintance Buys Your Data?:
  A New Privacy Harm in the Age of Data Brokers}.
\newblock \bibinfo{journal}{\emph{Yale J. Regulat.}} \bibinfo{volume}{34},
  \bibinfo{number}{2} (\bibinfo{year}{2017}), \bibinfo{pages}{667--707}.
\newblock
\urldef\tempurl%
\url{https://doi.org/10.2139/ssrn.2870044}
\showDOI{\tempurl}


\bibitem[\protect\citeauthoryear{Rudin}{Rudin}{2019}]%
        {Rudin}
\bibfield{author}{\bibinfo{person}{Cynthia Rudin}.}
  \bibinfo{year}{2019}\natexlab{}.
\newblock \showarticletitle{Stop explaining black box machine learning models
  for high stakes decisions and use interpretable models instead}.
\newblock \bibinfo{journal}{\emph{Nat. Mach. Intell.}}  \bibinfo{volume}{1}
  (\bibinfo{year}{2019}), \bibinfo{pages}{206--15}.
\newblock
\urldef\tempurl%
\url{https://doi.org/10.1038/s42256-019-0048-x}
\showDOI{\tempurl}


\bibitem[\protect\citeauthoryear{Russel}{Russel}{1997}]%
        {russel}
\bibfield{author}{\bibinfo{person}{S. Russel}.}
  \bibinfo{year}{1997}\natexlab{}.
\newblock \showarticletitle{Rationality and Intelligence}.
\newblock \bibinfo{journal}{\emph{Springer, Found. Rat. Agency}}
  (\bibinfo{year}{1997}).
\newblock


\bibitem[\protect\citeauthoryear{Sandler et~al\mbox{.}}{Sandler
  et~al\mbox{.}}{2019}]%
        {Accenture_AI_Committee}
\bibfield{author}{\bibinfo{person}{R. Sandler} {et~al\mbox{.}}}
  \bibinfo{year}{2019}\natexlab{}.
\newblock \showarticletitle{Building Data \& AI Ethics Comm'ts}.
\newblock \bibinfo{journal}{\emph{Accenture}} (\bibinfo{year}{2019}).
\newblock


\bibitem[\protect\citeauthoryear{Schiff, Biddle, Borenstein, and Laas}{Schiff
  et~al\mbox{.}}{2020}]%
        {Guides}
\bibfield{author}{\bibinfo{person}{Daniel Schiff}, \bibinfo{person}{Justin
  Biddle}, \bibinfo{person}{Jason Borenstein}, {and} \bibinfo{person}{Kelly
  Laas}.} \bibinfo{year}{2020}\natexlab{}.
\newblock \showarticletitle{What's Next for AI Ethics, Policy, and
  Governance?}. In \bibinfo{booktitle}{\emph{Proceedings of the AAAI/ACM
  Conference on AI, Ethics, and Society}} (New York)
  \emph{(\bibinfo{series}{AIES '20})}. \bibinfo{pages}{153--158}.
\newblock
\urldef\tempurl%
\url{https://doi.org/10.1145/3375627.3375804}
\showDOI{\tempurl}


\bibitem[\protect\citeauthoryear{SEC}{SEC}{2019}]%
        {SEC}
\bibfield{author}{\bibinfo{person}{SEC}.} \bibinfo{year}{2019}\natexlab{}.
\newblock \bibinfo{title}{SEC Proposes to Modernize Disclosures Under
  Regulation S-K}.
\newblock
\newblock


\bibitem[\protect\citeauthoryear{Selbst, Boyd, Friedler, Venkatasubramanian,
  and Vertesi}{Selbst et~al\mbox{.}}{2019}]%
        {selbst}
\bibfield{author}{\bibinfo{person}{Andrew~D. Selbst}, \bibinfo{person}{Danah
  Boyd}, \bibinfo{person}{Sorelle~A. Friedler}, \bibinfo{person}{Suresh
  Venkatasubramanian}, {and} \bibinfo{person}{Janet Vertesi}.}
  \bibinfo{year}{2019}\natexlab{}.
\newblock \showarticletitle{Fairness and Abstraction in Sociotechnical
  Systems}. In \bibinfo{booktitle}{\emph{Proceedings of the Conference on
  Fairness, Accountability, and Transparency}} (Atlanta, GA)
  \emph{(\bibinfo{series}{FAT* '19})}. \bibinfo{pages}{59--68}.
\newblock
\urldef\tempurl%
\url{https://doi.org/10.1145/3287560.3287598}
\showDOI{\tempurl}


\bibitem[\protect\citeauthoryear{Simonite}{Simonite}{2015}]%
        {google}
\bibfield{author}{\bibinfo{person}{Tom Simonite}.}
  \bibinfo{year}{2015}\natexlab{}.
\newblock \showarticletitle{Probing the Dark Side of Google's Ad-Targeting
  Systems}.
\newblock \bibinfo{journal}{\emph{MIT Tech. Rev.}} (\bibinfo{year}{2015}).
\newblock


\bibitem[\protect\citeauthoryear{Strickland}{Strickland}{2019}]%
        {ieee}
\bibfield{author}{\bibinfo{person}{Eliza Strickland}.}
  \bibinfo{year}{2019}\natexlab{}.
\newblock \showarticletitle{AI Agents Startle Researchers With Unexpected
  Hide-and-Seek Strategies}.
\newblock \bibinfo{journal}{\emph{IEEE Spectrum}} (\bibinfo{year}{2019}).
\newblock


\bibitem[\protect\citeauthoryear{Sun, Gaut, Tang, Huang, ElSherief, Zhao,
  Mirza, Belding, Chang, and Wang}{Sun et~al\mbox{.}}{2019}]%
        {NLP_bias}
\bibfield{author}{\bibinfo{person}{Tony Sun}, \bibinfo{person}{Andrew Gaut},
  \bibinfo{person}{Shirlyn Tang}, \bibinfo{person}{Yuxin Huang},
  \bibinfo{person}{Mai ElSherief}, \bibinfo{person}{Jieyu Zhao},
  \bibinfo{person}{Diba Mirza}, \bibinfo{person}{Elizabeth Belding},
  \bibinfo{person}{Kai-Wei Chang}, {and} \bibinfo{person}{William~Yang Wang}.}
  \bibinfo{year}{2019}\natexlab{}.
\newblock \showarticletitle{Mitigating Gender Bias in Natural Language
  Processing}. In \bibinfo{booktitle}{\emph{Proceedings of the 57th Annual
  Meeting of the Association for Computational Linguistics}}.
  \bibinfo{publisher}{ACL}, \bibinfo{address}{Florence, Italy},
  \bibinfo{pages}{1630--40}.
\newblock
\urldef\tempurl%
\url{https://doi.org/10.18653/v1/P19-1159}
\showDOI{\tempurl}


\bibitem[\protect\citeauthoryear{Szegedy, Zaremba, Sutskever, Bruna, Erhan,
  Goodfellow, and Fergus}{Szegedy et~al\mbox{.}}{2014}]%
        {Szegedy}
\bibfield{author}{\bibinfo{person}{Christian Szegedy},
  \bibinfo{person}{Wojciech Zaremba}, \bibinfo{person}{Ilya Sutskever},
  \bibinfo{person}{Joan Bruna}, \bibinfo{person}{Dumitru Erhan},
  \bibinfo{person}{Ian Goodfellow}, {and} \bibinfo{person}{Rob Fergus}.}
  \bibinfo{year}{2014}\natexlab{}.
\newblock \showarticletitle{Intriguing properties of neural networks}. In
  \bibinfo{booktitle}{\emph{ICLR '14}}. \bibinfo{numpages}{10}~pages.
\newblock
\showeprint[arxiv]{1312.6199}


\bibitem[\protect\citeauthoryear{Trippi and Lee}{Trippi and Lee}{1995}]%
        {portfolio}
\bibfield{author}{\bibinfo{person}{Robert~R. Trippi} {and}
  \bibinfo{person}{Jae~K. Lee}.} \bibinfo{year}{1995}\natexlab{}.
\newblock \bibinfo{booktitle}{\emph{AI in Finance and Investing}}.
\newblock \bibinfo{publisher}{Irwin Professional Publishing},
  \bibinfo{address}{Burr Ridge, IL}.
\newblock


\bibitem[\protect\citeauthoryear{Tsirtsis and Gomez~Rodriguez}{Tsirtsis and
  Gomez~Rodriguez}{2020}]%
        {tsirtsis2020decisions}
\bibfield{author}{\bibinfo{person}{Stratis Tsirtsis} {and}
  \bibinfo{person}{Manuel Gomez~Rodriguez}.} \bibinfo{year}{2020}\natexlab{}.
\newblock \showarticletitle{Decisions, Counterfactual Explanations and
  Strategic Behavior}.
\newblock   \bibinfo{volume}{33} (\bibinfo{year}{2020}),
  \bibinfo{pages}{16749--16760}.
\newblock


\bibitem[\protect\citeauthoryear{Tutt}{Tutt}{2017}]%
        {FDA}
\bibfield{author}{\bibinfo{person}{Andrew Tutt}.}
  \bibinfo{year}{2017}\natexlab{}.
\newblock \showarticletitle{An FDA for Algorithms}.
\newblock \bibinfo{journal}{\emph{Adm. Law Rev.}} \bibinfo{volume}{69},
  \bibinfo{number}{1} (\bibinfo{year}{2017}), \bibinfo{pages}{83--123}.
\newblock


\bibitem[\protect\citeauthoryear{Vanderelst and Winfield}{Vanderelst and
  Winfield}{2018a}]%
        {ethicslayer}
\bibfield{author}{\bibinfo{person}{Dieter Vanderelst} {and}
  \bibinfo{person}{Alan Winfield}.} \bibinfo{year}{2018}\natexlab{a}.
\newblock \showarticletitle{An architecture for ethical robots inspired by the
  simulation theory of cognition}.
\newblock \bibinfo{journal}{\emph{Cognit. Sys. Res.}}  \bibinfo{volume}{48}
  (\bibinfo{year}{2018}).
\newblock
\urldef\tempurl%
\url{https://doi.org/10.1016/j.cogsys.2017.04.002}
\showDOI{\tempurl}


\bibitem[\protect\citeauthoryear{Vanderelst and Winfield}{Vanderelst and
  Winfield}{2018b}]%
        {darkrobot}
\bibfield{author}{\bibinfo{person}{Dieter Vanderelst} {and}
  \bibinfo{person}{Alan Winfield}.} \bibinfo{year}{2018}\natexlab{b}.
\newblock \showarticletitle{The Dark Side of Ethical Robots}. In
  \bibinfo{booktitle}{\emph{Proceedings of the 2018 AAAI/ACM Conference on AI,
  Ethics, and Society}} (New Orleans, LA) \emph{(\bibinfo{series}{AIES '18})}.
  \bibinfo{pages}{317--322}.
\newblock
\urldef\tempurl%
\url{https://doi.org/10.1145/3278721.3278726}
\showDOI{\tempurl}


\bibitem[\protect\citeauthoryear{Veale and Borgesius}{Veale and
  Borgesius}{2021}]%
        {Veale}
\bibfield{author}{\bibinfo{person}{Michael Veale} {and}
  \bibinfo{person}{Frederik~Zuiderveen Borgesius}.}
  \bibinfo{year}{2021}\natexlab{}.
\newblock \showarticletitle{Demystifying the Draft EU Artificial Intelligence
  Act}.
\newblock \bibinfo{journal}{\emph{Computer Law Review International}}
  \bibinfo{volume}{22}, \bibinfo{number}{4} (\bibinfo{year}{2021}),
  \bibinfo{pages}{97--112}.
\newblock
\urldef\tempurl%
\url{https://doi.org/10.9785/cri-2021-220402}
\showDOI{\tempurl}


\bibitem[\protect\citeauthoryear{V{\'e}liz}{V{\'e}liz}{2019}]%
        {Veliz19}
\bibfield{author}{\bibinfo{person}{Carissa V{\'e}liz}.}
  \bibinfo{year}{2019}\natexlab{}.
\newblock \showarticletitle{Three things digital ethics can learn from medical
  ethics}.
\newblock \bibinfo{journal}{\emph{Nat. Electron.}} \bibinfo{volume}{2},
  \bibinfo{number}{8} (\bibinfo{year}{2019}), \bibinfo{pages}{316--8}.
\newblock
\urldef\tempurl%
\url{https://doi.org/10.1038/s41928-019-0294-2}
\showDOI{\tempurl}


\bibitem[\protect\citeauthoryear{Wachter}{Wachter}{2021}]%
        {Wachter}
\bibfield{author}{\bibinfo{person}{Sandra Wachter}.}
  \bibinfo{year}{2021}\natexlab{}.
\newblock \showarticletitle{Affinity Profiling and Discrimination By
  Association in Online Behavioral Advertising}.
\newblock \bibinfo{journal}{\emph{Berkeley Tech. Law J.}} \bibinfo{volume}{35},
  \bibinfo{number}{2} (\bibinfo{year}{2021}), \bibinfo{pages}{367--430}.
\newblock
\urldef\tempurl%
\url{https://doi.org/10.15779/Z38JS9H82M}
\showDOI{\tempurl}


\bibitem[\protect\citeauthoryear{Wakefield}{Wakefield}{2019}]%
        {BBC}
\bibfield{author}{\bibinfo{person}{Jane Wakefield}.}
  \bibinfo{year}{2019}\natexlab{}.
\newblock \showarticletitle{Google's Ethics Board Shut Down}.
\newblock \bibinfo{journal}{\emph{BBC News}} (\bibinfo{year}{2019}).
\newblock


\bibitem[\protect\citeauthoryear{{Wall St. J.}}{{Wall St. J.}}{2019}]%
        {wsj2}
\bibfield{author}{\bibinfo{person}{{Wall St. J.}}}
  \bibinfo{year}{2019}\natexlab{}.
\newblock \showarticletitle{What Your Face May Tell Lenders about
  Creditworthiness}.
\newblock  (\bibinfo{year}{2019}).
\newblock


\bibitem[\protect\citeauthoryear{Wallach and Allen}{Wallach and Allen}{2008}]%
        {wallach}
\bibfield{author}{\bibinfo{person}{Wendell Wallach} {and}
  \bibinfo{person}{Colin Allen}.} \bibinfo{year}{2008}\natexlab{}.
\newblock \bibinfo{booktitle}{\emph{Moral Machines}}.
\newblock \bibinfo{publisher}{Oxford}.
\newblock


\bibitem[\protect\citeauthoryear{Wallach, Allen, and Smit}{Wallach
  et~al\mbox{.}}{2008}]%
        {bottomup}
\bibfield{author}{\bibinfo{person}{Wendell Wallach}, \bibinfo{person}{Colin
  Allen}, {and} \bibinfo{person}{Iva Smit}.} \bibinfo{year}{2008}\natexlab{}.
\newblock \showarticletitle{Machine morality}.
\newblock \bibinfo{journal}{\emph{AI Soc.}} \bibinfo{volume}{22},
  \bibinfo{number}{4} (\bibinfo{year}{2008}), \bibinfo{pages}{565--82}.
\newblock
\urldef\tempurl%
\url{https://doi.org/10.1007/s00146-007-0099-0}
\showDOI{\tempurl}


\bibitem[\protect\citeauthoryear{Wexler, Pushkarna, Bolukbasi, Wattenberg,
  Vi{\'e}gas, and Wilson}{Wexler et~al\mbox{.}}{2020}]%
        {whatif}
\bibfield{author}{\bibinfo{person}{James Wexler}, \bibinfo{person}{Mahima
  Pushkarna}, \bibinfo{person}{Tolga Bolukbasi}, \bibinfo{person}{Martin
  Wattenberg}, \bibinfo{person}{Fernanda Vi{\'e}gas}, {and}
  \bibinfo{person}{Jimbo Wilson}.} \bibinfo{year}{2020}\natexlab{}.
\newblock \showarticletitle{The What-If Tool}.
\newblock \bibinfo{journal}{\emph{IEEE Trans. Vis. Comp. Graph.}}
  \bibinfo{volume}{26}, \bibinfo{number}{1} (\bibinfo{year}{2020}),
  \bibinfo{pages}{56--65}.
\newblock
\urldef\tempurl%
\url{https://doi.org/10.1109/TVCG.2019.2934619}
\showDOI{\tempurl}


\bibitem[\protect\citeauthoryear{{Wharton}}{{Wharton}}{2018}]%
        {penn}
\bibfield{author}{\bibinfo{person}{{Wharton}}.}
  \bibinfo{year}{2018}\natexlab{}.
\newblock \bibinfo{title}{How Fintech Serves the `Invisible Prime' Borrower}.
\newblock
\newblock


\bibitem[\protect\citeauthoryear{Winfield and Jirotka}{Winfield and
  Jirotka}{2018}]%
        {aireg}
\bibfield{author}{\bibinfo{person}{Alan F.~T. Winfield} {and}
  \bibinfo{person}{Marina Jirotka}.} \bibinfo{year}{2018}\natexlab{}.
\newblock \showarticletitle{Ethical governance is essential to building trust
  in robotics and artificial intelligence systems}.
\newblock \bibinfo{journal}{\emph{Phil. Trans. Roy. Soc. A}}
  \bibinfo{volume}{376}, \bibinfo{number}{20180085} (\bibinfo{year}{2018}),
  \bibinfo{numpages}{13}~pages.
\newblock
\urldef\tempurl%
\url{https://doi.org/10.1098/rsta.2018.0085}
\showDOI{\tempurl}


\bibitem[\protect\citeauthoryear{Witkowski}{Witkowski}{2019}]%
        {FDIC}
\bibfield{author}{\bibinfo{person}{Rachel Witkowski}.}
  \bibinfo{year}{2019}\natexlab{}.
\newblock \showarticletitle{Regulators must issue AI guidance or FDIC will}.
\newblock \bibinfo{journal}{\emph{Amer. Banker}} (\bibinfo{year}{2019}).
\newblock


\bibitem[\protect\citeauthoryear{{World Economic Forum}}{{World Economic
  Forum}}{2019}]%
        {WEF}
\bibfield{author}{\bibinfo{person}{{World Economic Forum}}.}
  \bibinfo{year}{2019}\natexlab{}.
\newblock \bibinfo{title}{Shaping Future of AI Governance}.
\newblock
\newblock


\bibitem[\protect\citeauthoryear{{World Economic Forum}}{{World Economic
  Forum}}{2020}]%
        {wef2020}
\bibfield{author}{\bibinfo{person}{{World Economic Forum}}.}
  \bibinfo{year}{2020}\natexlab{}.
\newblock \bibinfo{title}{Transforming Paradigms}.
\newblock
\newblock


\bibitem[\protect\citeauthoryear{Xiong, Liu, Zhong, Yang, and Walid}{Xiong
  et~al\mbox{.}}{2018}]%
        {XiongLZY2018}
\bibfield{author}{\bibinfo{person}{Zhuoran Xiong}, \bibinfo{person}{Xiao-Yang
  Liu}, \bibinfo{person}{Shan Zhong}, \bibinfo{person}{Hongyang Yang}, {and}
  \bibinfo{person}{Anwar Walid}.} \bibinfo{year}{2018}\natexlab{}.
\newblock \bibinfo{title}{Practical Deep Reinforcement Learning Approach for
  Stock Trading}.
\newblock
\newblock
\showeprint[arxiv]{1811.07522}~[cs.LG]


\bibitem[\protect\citeauthoryear{Yampolskiy}{Yampolskiy}{2016}]%
        {DangersAI}
\bibfield{author}{\bibinfo{person}{Roman~V. Yampolskiy}.}
  \bibinfo{year}{2016}\natexlab{}.
\newblock \showarticletitle{Taxonomy of Pathways to Dangerous AI}. In
  \bibinfo{booktitle}{\emph{2nd International. Workshop on AI, Ethics and
  Society (AIEthicsSociety2016)}}. \bibinfo{pages}{143--8}.
\newblock


\bibitem[\protect\citeauthoryear{Yampolskiy}{Yampolskiy}{2020}]%
        {unpredictableAI}
\bibfield{author}{\bibinfo{person}{Roman~V. Yampolskiy}.}
  \bibinfo{year}{2020}\natexlab{}.
\newblock \showarticletitle{Unpredictability of AI}.
\newblock \bibinfo{journal}{\emph{J. AI Conscious.}} \bibinfo{volume}{7},
  \bibinfo{number}{1} (\bibinfo{year}{2020}), \bibinfo{pages}{109--18}.
\newblock
\urldef\tempurl%
\url{https://doi.org/10.1142/S2705078520500034}
\showDOI{\tempurl}


\bibitem[\protect\citeauthoryear{Young, Rodriguez, Keller, Sun, Sa,
  Whittington, and Howe}{Young et~al\mbox{.}}{2019}]%
        {Open}
\bibfield{author}{\bibinfo{person}{Meg Young}, \bibinfo{person}{Luke
  Rodriguez}, \bibinfo{person}{Emily Keller}, \bibinfo{person}{Feiyang Sun},
  \bibinfo{person}{Boyang Sa}, \bibinfo{person}{Jan Whittington}, {and}
  \bibinfo{person}{Bill Howe}.} \bibinfo{year}{2019}\natexlab{}.
\newblock \showarticletitle{Beyond Open vs.\ Closed}.
\newblock  (\bibinfo{year}{2019}), \bibinfo{pages}{191--200}.
\newblock
\urldef\tempurl%
\url{https://doi.org/10.1145/3287560.3287577}
\showDOI{\tempurl}


\bibitem[\protect\citeauthoryear{Z\"{u}gner, Borchert, Akbarnejad, and
  G\"{u}nnemann}{Z\"{u}gner et~al\mbox{.}}{2020}]%
        {KDD}
\bibfield{author}{\bibinfo{person}{Daniel Z\"{u}gner}, \bibinfo{person}{Oliver
  Borchert}, \bibinfo{person}{Amir Akbarnejad}, {and} \bibinfo{person}{Stephan
  G\"{u}nnemann}.} \bibinfo{year}{2020}\natexlab{}.
\newblock \showarticletitle{Adversarial Attacks on Graph Neural Networks}.
\newblock \bibinfo{journal}{\emph{ACM Trans. Knowl. Discov. Data}}
  \bibinfo{volume}{14}, \bibinfo{number}{5}, Article \bibinfo{articleno}{57}
  (\bibinfo{year}{2020}), \bibinfo{numpages}{31}~pages.
\newblock
\urldef\tempurl%
\url{https://doi.org/10.1145/3394520}
\showDOI{\tempurl}


\end{thebibliography}
\bibliographystyle{ACM-Reference-Format}
\end{document}